\documentclass[14pt]{article}
\pdfoutput=1
\usepackage{amsmath,amssymb,amsfonts,amsthm,bm,bbm,cancel,wasysym}
\usepackage{epsfig,graphics,graphicx,epstopdf}
\usepackage{array,booktabs,colortbl,colordvi,multirow}
\usepackage{colordvi,color,xcolor}
\usepackage{hyperref}
\usepackage{rotating}

\usepackage{verbatim}
\usepackage{cite}
\usepackage{subfig}
\usepackage{setspace}
\usepackage{url}
\usepackage[percent]{overpic}
\usepackage{slashed}
\usepackage{authblk}
\usepackage{xspace}
\usepackage{fullpage}
\usepackage{hyperref}

\def\ie{{\it i.e.}}
\def\eg{{\it e.g.}}
\def\etc{{\it etc}}

\def\to{\rightarrow}

\newskip\zatskip \zatskip=0pt plus0pt minus0pt
\def\matth{\mathsurround=0pt}
\def\lsim{\mathrel{\mathpalette\atversim<}}
\def\gsim{\mathrel{\mathpalette\atversim>}}
\def\atversim#1#2{\lower0.7ex\vbox{\baselineskip\zatskip\lineskip\zatskip
  \lineskiplimit 0pt\ialign{$\matth#1\hfil##\hfil$\crcr#2\crcr\sim\crcr}}}

\begin{document}

%----------------------------------- TITLE AND AUTHORS -----------------------------------------%

%Preprint numbers
\begin{flushright}
SLAC-PUB-17628\\
\today
\end{flushright}
\vspace*{5mm}

\renewcommand{\thefootnote}{\fnsymbol{footnote}}
\setcounter{footnote}{1}

\begin{center}

{\Large {\bf Towards a UV-Model of Kinetic Mixing and Portal Matter II: Exploring Unification in an $SU(N)$ Group}}\\
%\vspace*{0.15cm}

\vspace*{0.75cm}

{\bf Thomas G. Rizzo}~\footnote{rizzo@slac.stanford.edu}

\vspace{0.5cm}

{SLAC National Accelerator Laboratory}\ 
{2575 Sand Hill Rd., Menlo Park, CA, 94025 USA}

\end{center}
\vspace{.5cm}

%--------------------------------------------- ABSTRACT ---------------------------------------------%

\begin{abstract}
\noindent

If dark matter (DM) interacts with the Standard Model (SM) via the $U(1)_D$ kinetic mixing (KM) portal at low energies, it necessitates not only the existence of portal matter (PM) particles 
which carry both dark and SM quantum numbers, but also a possible UV completion into which this $U(1)_D$ and the SM are both embedded. In earlier work, following a bottom-up approach, 
we attempted to construct a more unified framework of these SM and dark sector interactions.  In this paper we will instead begin to explore, from the top-down, the possibility of the 
unification of these forces via the decomposition of a GUT-like group, $G\to G_{SM}\times G_{Dark}$, where $U(1)_D$ is now a low energy diagonal subgroup of $G_{Dark}$ and where the 
familiar $G_{SM}=SU(5)$ will play the role of a proxy for the conventional $SU(3)_c\times SU(2)_L\times U(1)_Y$ SM gauge group. In particular, for this study it will be assumed that 
$G=SU(N)$ with $N=6-10$.  Although not our main goal, models that also unify the three SM generational structure within this same general framework will also be examined. 
The possibilities are found to be quite highly constrained by our chosen set of model building requirements which are likely too strong when they are employed simultaneously to obtain 
a successful model framework.
\end{abstract}

\vspace{0.5cm}
\renewcommand{\thefootnote}{\arabic{footnote}}
\setcounter{footnote}{0}
\thispagestyle{empty}
\vfill
\newpage
\setcounter{page}{1}

%-------------------------------- DOCUMENT: INTRODUCTION ---------------------------------%

% 1 Introduction

\section{Introduction and Background Discussion}

The nature of dark matter (DM) and its possible non-gravitational interactions with the Standard Model (SM) remain leading questions in particle physics. Given the measurements of the DM 
relic density from Planck\cite{Aghanim:2018eyx}, it is more than likely that interactions of {\it some} kind must exist coupling the SM to DM and possibly DM to itself. How would these `fit in' with 
the known forces of the SM in a unified framework and how are they generated? Although these are not new questions and we may be surprised by the eventual answers, 
the well-studied DM candidates, such as the QCD axion\cite{Kawasaki:2013ae,Graham:2015ouw,Irastorza:2018dyq} and weakly interacting massive particles, \ie, 
WIMPS\cite{Arcadi:2017kky,Roszkowski:2017nbc}, continue to be hunted for without success over a wide range of fronts and their allowed parameter spaces continue to 
be eaten into as a result of the null searches by direct or indirect detection experiments as well as those at the LHC\cite{LHC,Aprile:2018dbl,Fermi-LAT:2016uux,Amole:2019fdf}.  The lack of any 
traditional signatures has inspired a vast effort in examining an ever growing set of DM candidates spanning wide ranges in both DM particle masses and the strength of their couplings to 
the SM\cite{Alexander:2016aln,Battaglieri:2017aum,Bertone:2018krk}. It has been found that DM may couple to the SM in many various ways and one very useful tool to classify these 
possible interactions is via both renormalizable (\ie, dimension $\leq 4$) or non-renormalizable (\ie, dimension $> 4$) `portals'. These portals posit not only the existence of DM itself but also 
a new set of fields which act as mediators between the SM and the, potentially complex, dark sector of which the DM itself is likely the lightest member. Of the many examples, one 
that has gotten much attention in the recent literature due 
to its flexibility is the renormalizable kinetic mixing(KM)/vector portal\cite{KM,vectorportal} based upon a new gauge interaction. This scenario can allow for the DM to reach its abundance 
via the familiar thermal mechanism\cite{Steigman:2015hda,Saikawa:2020swg} albeit for sub-GeV DM masses and employing new non-SM interactions that so far could have evaded detection. 

Such a scenario can be realized in many ways enjoying various levels of complexity. The simplest manifestation assumes only the existence of a new $U(1)_D$ gauge group, with a coupling 
$g_D$, under which the SM fields are singlets, carrying  no dark charges, \ie, $Q_D=0$ and with the associated new gauge boson termed the `dark photon' (DP) 
\cite{Fabbrichesi:2020wbt,Graham:2021ggy}. $U(1)_D$ is usually assumed to be broken at or below the $\sim$ few GeV scale so that both  the DM and DP have comparable masses. 
The symmetry breaking in this model usually occurs via the vev(s) of at least one dark Higgs field in analogy with the spontaneous symmetry breaking in the SM. Within such a setup the 
the interaction between the SM and the dark sector is generated via kinetic mixing (KM) at the 1-loop level between $U(1)_D$ and the SM $U(1)_Y$ gauge fields. Specifically, these 
gauge bosons experience KM through the action of a set of fields, usually being vector-like fermions (or complex scalars), here called Portal Matter 
(PM)\cite{Rizzo:2018vlb,Rueter:2019wdf,Kim:2019oyh,Rueter:2020qhf,Wojcik:2020wgm,Rizzo:2021lob,Rizzo:2022qan,Wojcik:2022rtk,Rizzo:2022jti},
that carry both SM and $U(1)_D$ dark charges. After redefinitions back to canonically normalized fields removes the effect 
of the KM and both the SM and $U(1)_D$ symmetries are broken, this KM leads to a coupling of the DP to SM fields of the form $\simeq e\epsilon Q_{em}$, the origin of the DP 
nomenclature. The strength of the KM generated by these 1-loop vacuum polarization-like graphs is then described by a single dimensionless parameter, $\epsilon$, usually constrained 
by phenomenology to lie very roughly in the $\sim 10^{-(3-4)}$ range given the DM/DP sub-GeV mass region that we are assuming.  In the conventional normalization\cite{KM,vectorportal}, 
with $c_w=\cos \theta_w$, $\epsilon$ is given by the sum
\begin{equation}
\epsilon =c_w \frac{g_D g_Y}{24\pi^2} \sum_i ~\eta_i \frac{Y_i}{2}  N_{c_i}Q_{D_i}~ ln \frac{m^2_i}{\mu^2}\,,
\end{equation}
with $g_{Y,D}$ being the $U(1)_{Y,D}$ gauge couplings and $m_i(Y_i,Q_{D_i}, N_{c_i})$ are the mass (hypercharge, dark charge, number of colors) of the $i^{th}$ PM field. Here, 
$\eta_i=1(1/2)$ if the PM is a chiral fermion (complex scalar) and the hypercharge is normalized so that the electric charge is given by $Q_{em}=T_{3L}+Y/2$. In a more UV-complete 
theory, such as we are interested in here, this same group theory requires that the sum (for fermions and scalars separately)
\begin{equation}
\sum_i ~\eta_i \frac{Y_i}{2} N_{c_i} Q_{D_i}=0\,,
\end{equation}
so that $\epsilon$ is both finite and, if the PM masses are known, also calculable.  

It is important to address the question of how DM, this new $U(1)_D$ gauge interaction and the various PM fields might fit together with the known SM particles and gauge forces into a more 
unified structure and, as a result, to also consider the possibility that some further more complex gauge structure(s) might be kinematically accessible to existing and planned 
colliders in the future. This is a 
natural extension to the program of Grand Unification begun long ago\cite{Georgi:1974sy,Georgi:1974yf}, now augmented by a dark sector with its own matter content and gauge forces. 
In principle, in addressing 
these questions, one may want to follow either a bottom-up or a top-down approach, both of which have been previously discussed, with the former method  
followed in our previous studies\cite{Rueter:2019wdf,Wojcik:2020wgm}; in the discussion below, we will make a first attempt at a top-down analysis from our perspective. 
Of course, one might ask the 
obvious question if there is any reason to believe that this simple $U(1)_D$ scenario may itself already provide some indirect evidence as to it being part of a larger gauge structure, perhaps 
even one that is not too far away in energy scale. The following short exercise may be somewhat indicative.

\begin{figure}
\centerline{\includegraphics[width=4.8in,angle=0]{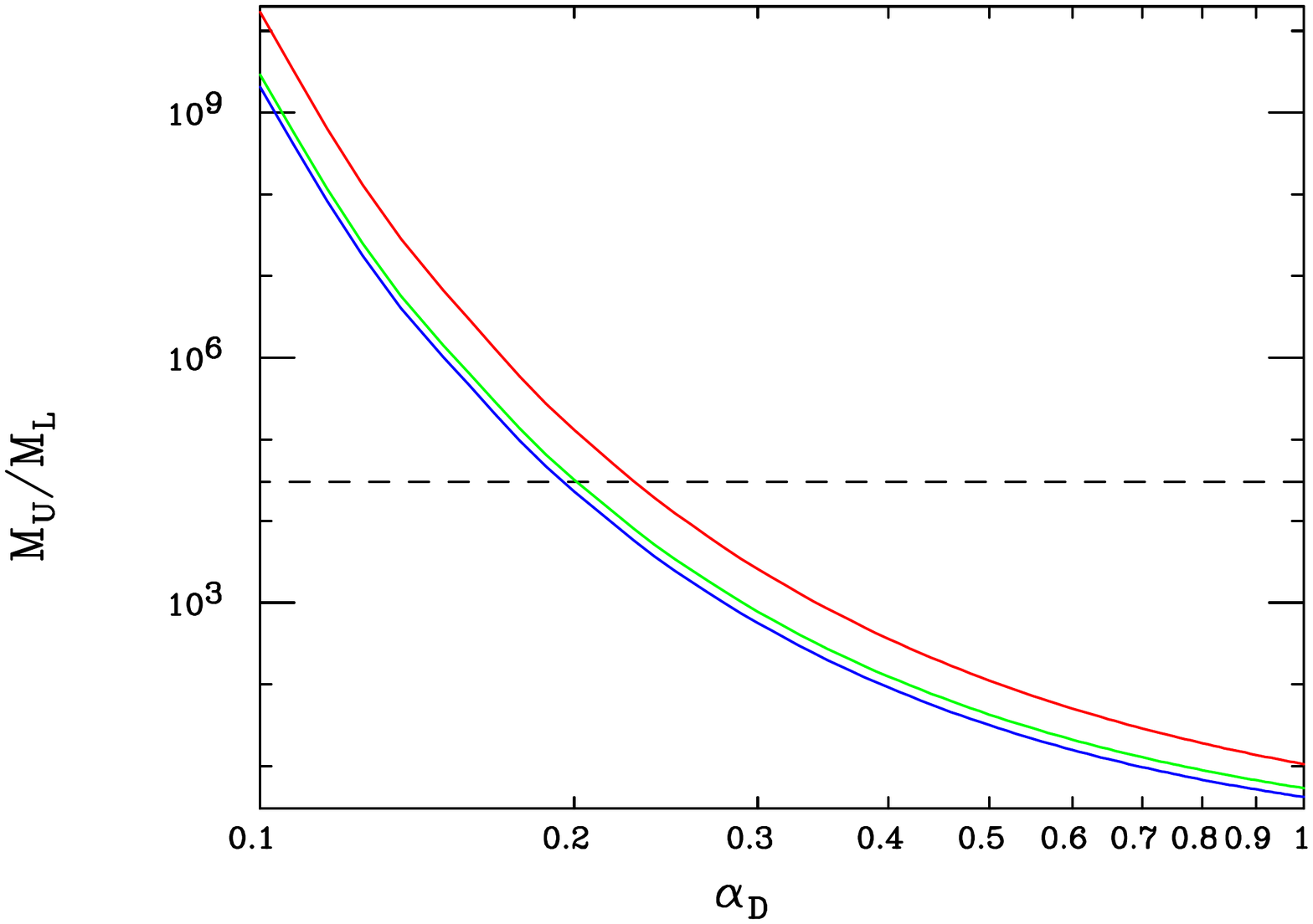}}
\vspace*{-2.3cm}
\centerline{\includegraphics[width=4.8in,angle=0]{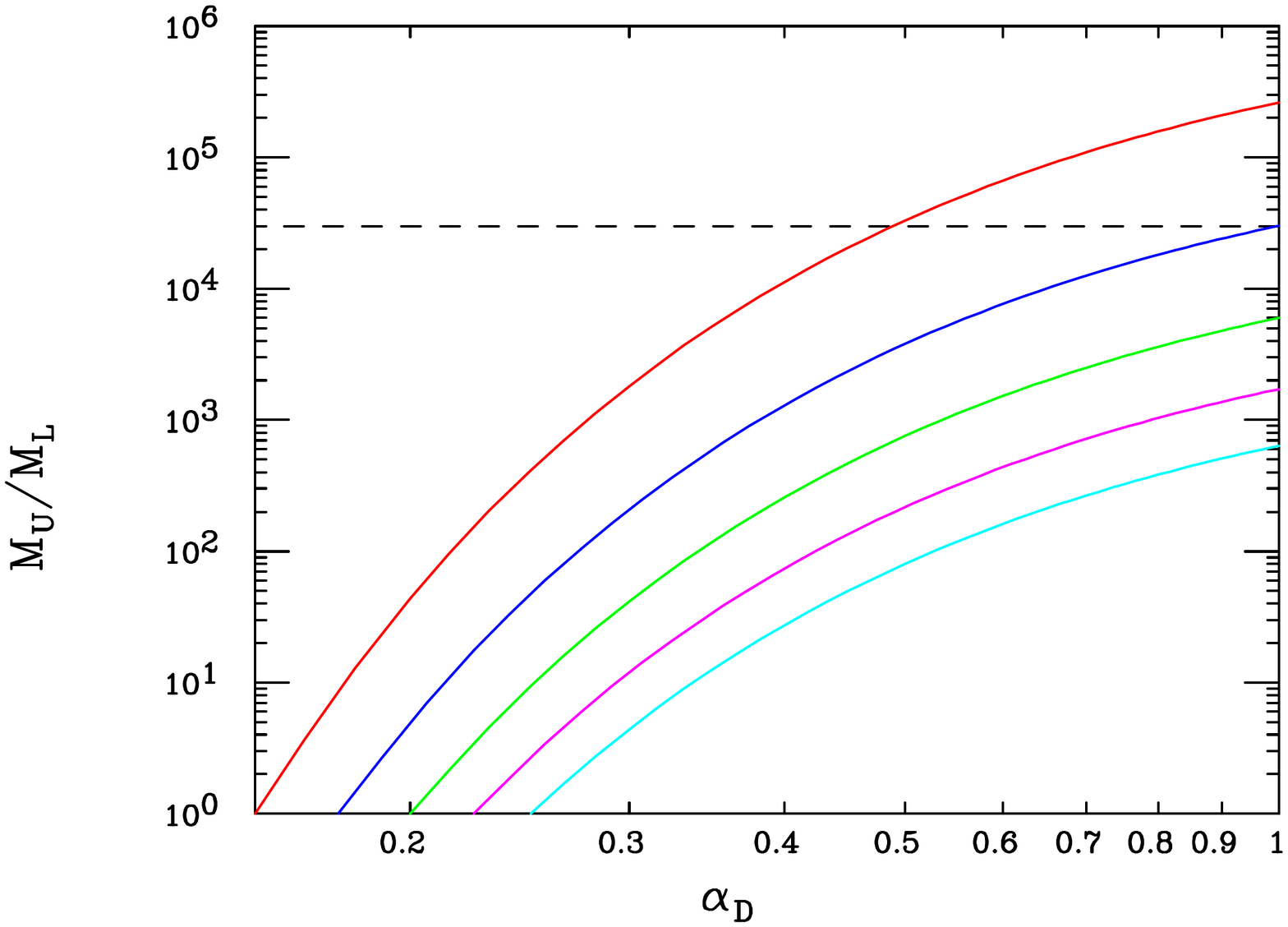}}
\vspace*{-1.30cm}
\caption{The running of $\alpha_D$ in the pseudo-Dirac DM example discussed in the text. The top panel shows the location of the Landau pole, $M_U$,  at the 1-loop (red), 2-loop (blue) and 
3-loop (green) level in RGE running, employing the $\bar{MS}$ scheme, as a function of $\alpha_D(M_L)$ where $M_L$ is the low energy scale associated with the DM, DP and dark 
Higgs fields, 
$\simeq 100$ MeV. The dashed line corresponds to $M_U=3$ TeV when $M_L=100$ MeV as a guide for the eye. The lower panel shows the 3-loop running of $\alpha_D$ (on the $x$-axis) 
in this same scenario as a function of $M_U/M_L$. The curves, from top to bottom, are for $\alpha_D(M_L)=0.15,0.175,0.20,0.225$ and 0.25, respectively. 
Here we observe that, \eg, if $\alpha_D(M_L=100$ MeV)$\geq 0.175$ then $\alpha_D \geq 1$ for $M_U \geq $ 3 TeV, again indicated by the dashed line. }
\label{fig1}
\end{figure}

At least for that part of the parameter space when the DM gauge coupling is somewhat large at low energies, it is useful consider the running of the coupling, \ie, $\alpha_D=g_D^2/4\pi$, into 
the UV.  As is very well-known, a $U(1)$ gauge theory is not asymptotically free and eventually will become strongly coupled or possibly experience a  Landau pole at some point as the energy 
scale increases. In a (more) UV-complete theory one would expect new physics of some form to enter before either of these things can happen so it is possible to roughly 
estimate by what energy scale this new dynamics must occur.  To be specific for demonstration purposes, consider the case of light fermionic DM, having $Q_D=1$, together with the 
DP and dark Higgs all lying in roughly similar mass 
range $M_L\sim 100$ MeV. More specifically, to escape the direct detection bounds due to elastic DM scattering as well as the rather strong constraints on $s$-wave annihilation from the 
CMB\cite{Aghanim:2018eyx,Slatyer:2015jla,Liu:2016cnk,Leane:2018kjk} for fermionic DM in this mass range, we consider the scenario where the DM is pseudo-Dirac with the relevant 
mass splitting between the two states generated by the same dark Higgs vev that is responsible for the mass of the DP.  Assuming that these fields are the only light degrees of freedom, 
one can run the value of 
$\alpha_D$ in a known manner from $M_L$ up to some higher scale, $M_U$ (or until some new physics with $Q_D\neq 0$ enters the RGE's) where one reaches a region of strong coupling 
and/or encounters a Landau pole{\footnote {The existence of other light fields with $Q_D\neq 0$ will only strengthen these arguments since then $\alpha_D$ will run even more quickly.}}. 
The SM fields do {\it not} enter into this calculation as they all have $Q_D=0$ and so will not couple to the DP to LO in the $\epsilon \to 0$ limit. The result of this simple calculation can be 
found in Fig.~\ref{fig1}. Here we can see that if $\alpha_D(M_L) \geq 0.175(0.20)$, a not infrequent assumption made in many phenomenological 
analyses\cite{Alexander:2016aln,Battaglieri:2017aum,Bertone:2018krk,Schuster:2021mlr}, its value will become non-perturbative (or even encounters a Landau pole\cite{Gockeler:1997dn}) 
before $M_U\simeq $ a few TeV when run up from the $M_L=100$ MeV scale. Even for $\alpha_D(M_L) \simeq 0.12$ (or 0.07), new physics must enter before the 
$\simeq 1000$ TeV (or the traditional grand unification) scale is reached. Semi-quantitatively, one finds that this conclusion is not very dependent as to whether these calculations 
are performed at the 1-, 2- or 3-loop level as can be gleaned from the Figure.  Although this simple toy example is only indicative, it gives support to the likelihood that a more complex, 
probably non-abelian, broken dark sector gauge structure\cite{Rueter:2019wdf,Wojcik:2020wgm,Murgui:2021eqf} is likely to be encountered at higher energies, and, perhaps in some 
cases, may 
not lie too far above the weak scale, perhaps even being accessible at, \eg, the HL-LHC and at other possible future colliders. Of course, in a top-down analysis, one may speculate that this 
transition is allowed to occur anywhere below the `unification' scale, $\sim 10^{16}$ GeV yet still above several TeV or so.

In what follows we will attempt a preliminary 
top-down construction wherein the SM gauge interactions and those of an enlarged dark sector are combined into single 'unification' group, $G$, which is 
broken at some very large scale to the product $G_{SM}\times G_{Dark}$ with the $U(1)_D$ totally contained within $G_{Dark}$. $G_{SM}$ certainly contains (at least) the usual SM 
$SU(3)_c\times SU(2)_L\times U(1)_Y$ (aka $3_c2_L1_Y$) subgroup. Perhaps the most {\it minimal} choice of a simple group is the the identification $G_{SM}=SU(5)$ so that, \eg, $G$ itself 
might naturally be further identified with $SU(N), ~N\geq 6$. We note that this choice of $G_{SM}$ is far from a unique one, even if we assume it to be simple, \ie, $SO(10)$ and 
$E_6$\cite{e6}  both certainly come to mind as does the Pati-Salam product group $SU(2)_L\times SU(2)_R\times SU(4)_c$\cite{Pati:1974yy} or even $[SU(3)]^3$\cite{Glashow:1984gc} in 
the case of non-simple groups.{\footnote{One can even imagine that $G$ is itself a product group, \eg, $SU(8)_L\times SU(8)_R$  or even just 
$SU(5)_A\times SU(5)_B$\cite{Dey:1995ar,Lonsdale:2014wwa} with $G_{Dark}$ embedded differently.}}  In the setups considered here, 
we expect that the heavy ($\gsim$ a few TeV) fermionic PM fields (which should be vector-like with respect to the SM) will transform non-trivially under both groups while dark matter and other 
purely dark sector fields will 
only transform non-trivially under $G_{Dark}$. No matter how $G_{Dark}$ itself gets broken, it will be required that $U(1)_D$ survives intact down to the $\sim 1$ GeV scale and some special 
`protection' along the way will usually be required to insure this remains the case. Numerous additional constraints will also need to be imposed in such a setup following from a set of (perhaps 
too strict) model building assumptions. Our goal will not be to consider in any detail the phenomenological implications of such models that are judged to be `successful' (if any) in this regard, 
but we will instead concentrate our efforts on the difficult task of attempting to satisfy all of these basic model building constraints. As will be fully discussed in the following Section, we 
will examine the corresponding unification group/gauge structures in some detail as well as the various steps of the symmetry breaking chain leading to the SM and allowing for the possibility 
of an unbroken $U(1)_D$ down to the $\lsim 1$ GeV scale. 

The outline of this paper is as follows: In Section 2, we provide the details of the basic model framework and the list of assumptions that we will be making in the analysis that follows. Many of 
these assumptions are fairly `traditional' ones and will be quite familiar from the extensive literature on the subject of grand unification including the SM family structure, some of it dating back 
well over 40 years. Others, will however, be guided by the additional requirements we need to impose on the dark sector to generate the appropriate PM masses while simultaneously 
maintaining an unbroken $U(1)_D$ down to the low energies below the electroweak scale.  In Section 3, we will systematically analyze in detail some representative examples of the set of 
$G=SU(N)$ models, for the $N=6-10$ series of scenarios with increasing $N$, from which, as $N$ grows, we will learn valuable lessons that can be employed for even larger values 
of $N$. At the end of this Section, we also briefly consider 
values of $N>10$ as well as some simple alterations in our set of model building assumptions in light of the previous obtained results.  A discussion of our final 
results, some possible future avenues of further investigation and our conclusions are then presented in Section 4.

%-------------------------------- DOCUMENT: SECTIONS -----------------------------------------%

\section{Basic Framework and Analysis Assumptions}

To begin our search for viable candidate models, we need to set out the underlying assumptions that we'll be making in the analysis that follows. Many of these, in one form or another, are 
very familiar and well-known,  having been the pillars for many studies in the literature related to extended Grand Unified Theories (GUTS) and the family/generation problem for over 
four decades. We will make use of these studies here but for a different purpose, \ie, the incorporation of a dark sector with, \eg, PM fields and new interactions. In particular, 
to see how far we can get this way, we generally will follow the rather conventional, strictly renormalizable, 4-d~{\footnote {For example, we do not consider the possibility of using 
orbifolds/boundary conditions to break gauge symmetries\cite{boundary} in higher dimensional setups.}}, 
non-supersymmetric approach to these constructions as described quite early on by Georgi\cite{Georgi:1979md}, but with some additions and modifications introduced almost 
immediately following this in  
the work of others authors, \eg, \cite{Frampton:1979cw,Frampton:1979fd,Frampton:1979tj,Vaughn:1979sm,Langacker:1980js,Kashibayashi:1982nu}.  
The further model building requirements and modifications discussed below are specifically inspired to account for the existence of (with respect to 
SM interactions) vector-like fermionic PM, which are charged under both the visible and dark sectors and that have masses above the electroweak scale, as well as the survival of an 
unbroken $U(1)_D$ down to very low energies $\lsim 1$ GeV~{\footnote {As is usual, all fermions will be are taken to be left-handed in the analysis below.}}. We note, however, the common 
presence of additional {\it scalar} PM with both SM and dark quantum numbers throughout these analyses as such fields are an integral part of the representations necessary for the 
breaking of the various symmetries that we will encounter. 

It is to be noted that some of our assumptions, certainly in combination, may be somewhat overly restrictive thus making it quite 
difficult for any model to pass through all the necessary hoops to be successful. The relative importance of the different constraints does change somewhat as $N$ increases as we will see. Be 
that as it may, our approach can allow us to identify where certain assumptions might be too strictly applied and so point us in future directions for model building. 

For our study below, the specific model building requirements will be taken to be as follows: 
\begin{enumerate}
\item {We will assume that the unifying group $G$ decomposes as $G_{SM}\times G_{Dark}$ with, for simplicity here, $SU(5)$ acting as a proxy for the SM and playing that role place of 
$G_{SM}=3_c2_L1_Y$. Thus $U(1)_D$ is, trivially, a diagonal subgroup of $G_{Dark}$ that remains unbroken down to the $\sim 1$ GeV scale so that the dark photon's coupling to the SM 
results from abelian KM with the SM $U(1)_Y$.  As noted above, the assignment of $SU(5)$ as the SM proxy is certainly far from unique and other choices, \eg, $SO(10)$, $E_6$,  \etc,  
are clearly possible.  
It is also possible that $G_{SM}$, though larger than $3_c2_L1_Y$, may not be a compact group; this is a strong model-building assumption, \ie, that the `pure SM' physics sector is 
itself not also extended by, \eg, additional $U(1)$ and/or $SU(2)$ factors. 
We further note that it is easily possible that $3_c2_L1_Y$ may be embedded into $G$ in quite a different manner and not in this simple product-like fashion as is assumed here.}

\item {Since the rank of $G\geq 5$ and must have complex irreducible representations, $[\bf {R_i}]$, but which are simultaneously required to be real with respect to $SU(3)_c$, it will 
be assumed 
that $G=SU(N)$. For simplicity and to obtain representations with (relatively) small dimensionality, it will be further assumed that the various irreducible representations that appear, 
$[\bf {R_i}]$, are solely obtained by taking antisymmetric products of the $SU(N)$ fundamental representation, $\bf {N}${~\footnote {This need not be the case but other options have been 
shown to lead to both representations and sets of such representations of significantly greater dimensionality hence more degrees of freedom; see for example, Ref.\cite{Feger:2015xqa}.}}. 
As is well-known, 
this assumption prevents the various resulting fermions from having non-SM-like $SU(3)_c$ and/or weak isospin transformation properties. Note that it will {\it not} be assumed that the 
fermion fields must all lie within a single irreducible representation. It is to be noted that this requirement not only restricts the set of SM $SU(5)$ representations that may appear 
but simultaneously {\it also} those of $G_{Dark}$ in a similar fashion and thus may be too strong of an assumption.}

\item {The combined set of all relevant irreducible fermionic representations $[\bf {R_i}]$ of $G=SU(N)$ under consideration for any of the models discussed here, taken together, 
are assumed to lead to $G$ being anomaly-free.}

\item {The $SU(N)$ gauge group above the `unification scale' will be assumed to have the property of asymptotically freedom (AF) at the one-loop level; here we will include in the relevant 
$\beta$-function the gauge and fermion contributions as well as the contributions from the minimal set of Higgs scalars required to break all of the various gauge symmetries and to 
generate the required particle masses. Specifically, we will demand that 
\begin{equation}
\beta_N=-\frac{11}{3} N+\frac{2}{3}\Sigma_f ~\eta^fT(R_f) + \frac{1}{3}\Sigma_s ~\eta^sT(R_s) <0 \,,  
\end{equation}
where the sums extend over the full set of complex (real) chiral fermion representations with $\eta^f=1(1/2)$  and, correspondingly, complex(real) scalar representations with 
$\eta^s=1(1/2)$.
As is perhaps obvious, we will see, as $N$ increases, that the overall dimensionality and corresponding $\beta$-function contributions of the various fermion and scalar $[\bf{R_i}]$ do as well 
(as does the overall number of degrees of freedom), making this condition ever more difficult to satisfy for large $N$. Note that we will {\it not} make the additional, potentially strong, assumption 
of requiring asymptotic freedom for the usual QCD $\beta$-function itself near/between the PM and unification scales. In some sense, one may wonder if this AF assumption is really 
justifiable\cite{safe,Christensen:2005bt}. We can also imagine many different effects, \eg, gravitational influences\cite{Hill:1983xh}, that may become quite important significantly 
above the unification scale, especially, as we will 
see that the role of AF becomes quite important in the discussion below.  Note that the introduction of supersymmetry would only result in a strengthening of the AF condition. The runnings of the 
individual gauge and Yukawa couplings below the unification scale for the models passing all of 
our requirements (if any) will not be examined here and is left for future work.}

\item {We will specifically assume the $SU(N)\to SU(5)\times SU(N-5)'\times U(1)_N$ breaking decomposition via the $SU(N)$ adjoint representation so that we can identify 
$G_{Dark}=SU(N-5)'\times U(1)_N$. {\footnote {Note that we are not allowing for the possibility that $G_{Dark}$ could just be a smaller subgroup of $SU(N-5)'\times U(1)_N$.}} 
In such a case, the fundamental representation $\bf N$ decomposes as $\bf {N} \to \bf {(5,1)}_{N-5}+\bf {(1,N-5)}_{-5}$. Since the group algebra for the 
anti-symmetric products of the $\bf 5$ of $SU(5)$ closes rapidly, one finds that only fields transforming under the SM $SU(5)$ as $\bf {1,5,10}$ and/or their complex conjugates 
will appear in the set $[\bf {R_i}]$, \ie, this set of representations can be symbolically 
decomposed under $SU(5)$ as\cite{Georgi:1979md} 
\begin{equation}
[\bf {R_i}] \to n_1 (\bf {1}) + n_5 (\bf {5}) + n_{10} (\bf {10}) + n_{\bar{10}} (\bf {\bar {10}}) +n_{\bar{5}} (\bf {\bar 5})\,,  
\end{equation}
under which the number of SM generations, $n_g$, which consists of a single $\bf {\bar 5}+ \bf {10}$ of $SU(5)$, is given by the difference\cite{Georgi:1979md} 
\begin{equation}
n_g=n_{\bar{5}} -n_5= n_{10}-n_{\bar {10}} \,.  
\end{equation}
Additional $SU(5)$ non-singlet fields beyond the three sets of $(\bf {\bar 5+10})$, which are to be identified with the usual SM fermions, might be identifiable as PM if they also satisfy other 
necessary requirements, \eg, in the case of fermions, they must be vector-like with respect to the SM and carry $Q_D\neq 0$. Additional pure SM singlet fields are, of course, also allowed and  
will appear as potentially dark sector fields. Note that in all generality in this decomposition, $Q_D$ must be given by the sum of generators
\begin{equation}
Q_D=\sum_i a_i \lambda_i^{Diag} +bQ_N\,, 
\end{equation}
where the $a_i,b$ are constant coefficients,  $\lambda_i^{Diag}$ are the well-known set of $N-6$ diagonal generators of $SU(N-5)'$, \ie, $\lambda_{3,8,15,24,...}$, \etc. and $Q_N$ 
is the $U(1)_N$ charge.}

\item {While the SM fields must have $Q_D=0$ by construction, they need {\it not} be singlets under the full $G_{Dark}$ if the $U(1)_D$ is `properly' embedded within it. Similarly, PM fields 
must carry $Q_D\neq 0$ and must also transform non-trivilally under $G_{SM}=SU(5)$ since they are required to carry SM quantum numbers, in particular, have non-zero hypercharges, $Y$, 
to induce $U(1)_Y-U(1)_D$ abelian KM. As noted, the fermionic PM fields must also be vector-like with respect to the SM to avoid numerous well-known constraints from, \eg,  precision 
electroweak measurements, direct searches, unitarity bounds and Higgs coupling determinations. The masses of these fermionic PM states must lie above the electroweak scale and also 
likely $\gsim 1-2$ TeV\cite{Rizzo:2018vlb,Rueter:2019wdf,Kim:2019oyh,Rueter:2020qhf,Wojcik:2020wgm,Rizzo:2021lob,Rizzo:2022qan,Wojcik:2022rtk,Rizzo:2022jti}  
depending upon their electroweak and color transformation properties.  Note that only the $Q_D=0$ components of the various scalar representations acting as 
Higgs fields can obtain vevs that are larger than $\sim1$ GeV to enable the survival of the low energy KM scenario. Since the fermionic PM fields generally lie in various representations 
of $G_{Dark}$ and obtain their masses via the Higgs mechanism, they will be chiral with respect to at least some of the $G_{Dark}$ subgroups.}

\item {We will assume that the set of representations, $[\bf {R_i}]$, can lead to the three SM generations in various ways, the most simple being 3 copies of a smaller set of representations 
as is the case in ordinary $SU(5)$; we will refer to models in this class as having `$n_g$=1' and true `family unification' is absent in such scenarios.  A more complex and interesting 
possibility, which we'll refer to as `$n_g=3$', constructs the three SM generations in a manner that allows any given representation or representations, $[\bf R]$, to appear more than 
once but the full set of all representations is {\it not} a triplification of a smaller subset of fields; this is a clearly a much stronger demand than the previous one and, as we will see, 
will require representations of larger rank to make remotely workable. We remind the Reader that in 
Georgi's original work\cite{Georgi:1979md}, the even stronger requirement was made that {\it no} representation could appear more that once in the set $[\bf {R_i}]$ but this requirement 
was subsequently relaxed rather soon by other authors, \eg, \cite{Frampton:1979cw,Frampton:1979fd,Frampton:1979tj,Vaughn:1979sm,Langacker:1980js} with an eye towards reducing 
the total number of degrees of fermionic freedom and also easing the AF requirement. Here, we will follow these later authors and place some emphasis on models which have a smaller 
overall number of additional degrees of 
freedom beyond the usual ones of the SM although `simple' models of both classes will be investigated. Obviously, fermionic PM will also come in three generations in the $n_g=1$ models 
but this need not be, and will likely not be, the case when $n_g=3$. Note that family unification itself, as traditionally discussed, is it {\it not} a goal of the current study. }

\item{Higgs fields must be present to break $SU(5)$ and also give the SM fermions their masses as usual as well as to break $G_{Dark}$, possibly in stages, down to $U(1)_D$. 
The Higgs fields at the penultimate stage of $G_{Dark}$ symmetry breaking, which will also generally lead to the masses of a set of vector-like (with respect to the SM) fermions which 
we might directly identify with PM, with masses above the electroweak scale and must allow for the existence of an unbroken $U(1)$ that we can identify with the low energy $U(1)_D$ 
under which the SM fields are, by assumption, neutral. As we will see, our assumptions then lead unambiguously to the symmetry breaking chain 
\begin{equation}
~~G\to G_{SM}\times G_{Dark}, ~~ G_{Dark}\to .... \to SU(2)_D \to U(1)_D\,, 
\end{equation}
which will be discussed in much detail below where we will consider the algebraically simpler scenario where the breaking of $G_{Dark} \to SU(2)_D=SU(2)'$ happens in a single step; 
the possibility of multistep breaking will not alter the results obtained here in any essential manner although their will undoubtedly be numerical impacts on the RGE's of the various 
gauge couplings. Note that the $SU(2)_D\to U(1)_D$ breaking cannot occur via the fundamental doublet representation but via, \eg,  the adjoint triplet. The usual $U(1)_D$ can then 
eventually broken at the $\sim 1$ GeV scale or below by the vevs of the many possible $Q_D\neq 0$ neutral Higgs scalars that we'll encounter which may also be (but need not be) SM 
singlets. It is also possible that this $U(1)_D$ may be broken by the Stueckelberg mechanism\cite{Feldman:2007wj} but that will not be helpful in, \eg,  generating any needed DM mass 
terms nor will be it helpful in solving some of the other model building issues with the symmetry breaking chains that we will subsequently face as they involve non-abelian symmetry breakings.}

\item{As noted, we will limit our set of possible particle interactions to those which are renormalizable at each level of symmetry breaking, \ie, we will not consider the contributions of potential 
higher dimensional interactions/operators arising from integrating out possible heavy fields appearing in loops.}

\item{Although we will not employ it directly as a model building constraint {\it {per se}}, the nature of the DM in this class of models is of some relevance. As noted in the Introduction, 
for thermal DM in the mass range $\lsim 1$ GeV anticipated here, CMB constraints tell us that its annihilation must be substantially suppressed at later times to avoid too much of an injection 
of electromagnetic energy into the evolving plasma\cite{Aghanim:2018eyx,Slatyer:2015jla,Liu:2016cnk,Leane:2018kjk}. One way to to do this is to require that this process be $p$-wave so 
that it is becomes velocity-squared suppressed later on after freeze-out and such a situation is most easily realized when the DM is a SM singlet, $Q_D\neq 0$, complex scalar that does 
not obtain a vev. Although we will not make any specific identification of such a field and some fine-tunings of the scalar potential may be required, as we will see below the opportunities for 
the existence of such fields will be quite numerous as they will always occur (at the very least) in both the fundamental and the second rank antisymmetric Higgs representations of $SU(N)$ 
of which we will make frequent use. Note that for $N>6$ such fields will generally transform non-trivially under $G_{Dark}$. The interplay of this type of DM with the similarly light dark Higgs 
field(s) may itself be rather complex\cite{Rizzo:2022qan}.}

\end{enumerate}

These combined requirements, though individually quite reasonable, are together very highly (perhaps overly) constraining as we will now discover by looking at a broad set of examples. As 
we will see, in particular, the combined requirements of asymptotic freedom,  successful mass generation for all of the fermionic PM fields at or above the TeV scale while simultaneously also 
requiring electroweak scale masses for the three families of SM fermions and a $U(1)_D$ that survives unbroken down to the $\sim 1$ GeV mass range are extremely difficult to satisfy. 
Clearly, careful but intentional violations of any one or more of these model building assumptions will lead to broad avenues for possible future investigations.

\section{Model Survey}

We now turn to our search for candidate gauge groups with specific fermion representations that satisfy all of the criteria above with ever increasing values of $N$ - beginning, briefly, 
with the educational case of $N=6$.

\subsection{$SU(6)$}

We start our analysis by quickly considering the $SU(6)$ unification scenario as it provides a very simple toy example of where things go badly wrong almost from the start. 
We begin by employing the familiar $SU(6)\to SU(5)\times U(1)_6$ breaking pattern wherein a single SM generation is embedded in the anomaly-free set of representations 
$2(\bf {\bar 6}) + \bf {15}$ (which is just the $\bf{27}$ of $E_6$\cite{Hewett:1988xc}, which under $SU(5)\times U(1)_6$ is simply $2[\bf {\bar 5}_{-1}+\bf {1}_5]+[\bf {10}_2+\bf {5}_{-4}]$ in 
obvious notation. This example is educational because it contains both 
an additional set of vector-like (with respect to the SM) $\bf {5}+\bf {\bar 5}$ fermion fields which {\it could} play the role of PM as well as 
an additional abelian gauge group, $U(1)_6$, which {\it could} be just $U(1)_D$, two of the necessary ingredients we require for a successful model. However, we see immediately that this 
is not the case as this possibility fails in the most trivial way: all of the fermions representations are seen to carry a non-zero $U(1)_6$ charge while we have demanded that all of the ordinary SM 
chiral fields have $Q_D=0$ so that we {\it cannot} identify $U(1)_D$ with $U(1)_6$. Since $U(1)_6$ is the only new gauge group factor beyond $SU(5)$ this simple possibility is clearly excluded. 
Furthermore, in parallel with this, we note that requiring the SM fermions to have $Q_D=0$ would also force the potential candidate PM fields to also have $Q_D=0$ in this framework.  
Specifically, in $n_g=1(3)$ type scenarios, we require that we can identify one (three) $\bf{\bar 5+10}$ representations with the SM fermions and necessarily having $Q_D=0$. 

This $SU(6)$ discussion teaches us a 
valuable, if perhaps obvious, lesson when considering the more general $SU(N)\to SU(5)\times SU(N-5)'\times U(1)_N$ decomposition as we will see below. The set of relevant $SU(N)$ chiral 
fermions will very commonly include at least one $\bf{ N}$ (or its conjugate), which subsequently decomposes as noted above as $\bf {N} \to \bf {(5,1)}_{N-5}+\bf {(1,N-5)}_{-5}$ whose 
first contributor we will commonly want to identify as `the SM $\bf {5}$' of $SU(5)$. Since this field necessarily has a non-zero $U(1)_N$ charge and is also a singlet of $SU(N-5)'$ so that 
no other diagonal generators are relevant, we 
must conclude that $U(1)_D$ has {\it no} contribution from $U(1)_N$ in this type of construction. Similarly, except for accidental cases, this also implies that all of the representations obtained 
via antisymmetric products of $\bf N$ with itself will also carry non-zero values of $U(1)_N$. 
Together this directly implies that in such setup we must have 
\begin{equation}
Q_D=\sum_i a_i \lambda_i^{Diag} \,, 
\end{equation}
where the $\lambda_i^{Diag}$ are defined above and any potential $Q_N$ contribution to $Q_D$ must now be {\it absent}, \ie, $b=0$. The same argument applies in the presence of any of the 
Higgs representations that 
produce SM fermion masses. If these are singlets of $SU(N-5)'$ then either their values of $Q_N$ must all be zero or $Q_N$ cannot be allowed to contribute to $Q_D$ in such a setup otherwise 
$U(1)_D$ would be broken at the electroweak scale. This simple result has non-trivial implications and we will see how it will 
play out more clearly in the subsequent examples we analyze more fully below. An important exemption to this conclusion {\it may} occur in scenarios where the spectrum of states is sufficiently 
rich that we can try to identify all of the SM $\bf {\bar 5}$'s and $\bf{10}$'s with fields which are {\it not} also $SU(N-5)'$ singlets and, simultaneously,  all carry non-zero values of $Q_N$.  
Such very rare cases, however, will encounter other problems such as, \eg, running afoul of the $SU(N)$ $\beta$-function constraint above or having Higgs fields which are 
$SU(N-5)'$ singlets carrying a non-zero $Q_N$ charge.

Note that since almost all the Higgs fields that we will encounter below will carry a non-zero value for the $U(1)_N$ charge, this symmetry will generally be broken at the same mass scale 
where the $SU(N-5)'$ group itself first breaks.

\subsection{$SU(7)$}

$SU(7)$ offers another opportunity to see where our requirements will cause models to fail and the general setup again simply `goes wrong' although in ways which are a bit more subtle 
than in the $SU(6)$ 
example above. Note that in this case, since $G_{Dark}=SU(2)'\times U(1)_7$ (and some of the SM fields are always in $SU(2)'$ singlets), that $Q_D$ must be proportional to the 
diagonal $\lambda_3$ generator of $SU(2)'$ by the arguments made above. 

$SU(7)$ has been considered as a potential GUT/family group since the earliest days and we can take advantage of that huge body of work here. There are many sets of $SU(7)$ 
representations which satisfy (most of) our basic requirements that have been previously examined in other 
contexts, \eg, \cite{Georgi:1979md,Frampton:1979cw,Vaughn:1979sm,Kang:1981nr,Chkareuli:2000bm,Chen:2021ovs}
These models differ mainly in how they address the 
family/generation problem, \ie, in the notation introduced above, \ie, are they of the `$n_g=1$' or `$n_g=3$' variety? Certainly the former are somewhat simpler but both types of setups 
will lead to similar problems in the present context. The following are a non-exhaustive but fairly representative set of asymptotically-free 
scenarios of both model classes appearing in the literature\cite{Georgi:1979md,Frampton:1979cw,Vaughn:1979sm,Kang:1981nr,Chkareuli:2000bm,Chen:2021ovs}: 
 \begin{eqnarray}
 (a)&~~~~& 3[3(\bf {\bar 7})+\bf {21}]\nonumber\\
 (b)&~~~~& 3[2(\bf {\bar 7})+\bf {35}]\nonumber\\
 (c)&~~~~& 8(\bf{\bar 7}) +2(\bf {21})+\bf {35}\nonumber\\
 (d)&~~~~& 7(\bf{\bar 7}) +2(\bf {35})+\bf {21}\,,
 \end{eqnarray} 
with the first two of these being examples of scenarios with $n_g=1$ while the last two are examples of $n_g=3$~{\footnote{Note that a pair of singlet fields should also appear to remove  
Witten anomalies\cite{Chen:2021ovs} but are not directly relevant to our present discussion so they are omitted for simplicity.}}. Note that other scenarios can be easily constructed\cite{Vaughn:1979sm} by 
`adding/subtracting' multiples of the combination of representations $\bf{\bar 7}+\bf{21}+\bf{\bar{35}}$\cite{Eichten:1982pn,Fonseca:2015aoa} which forms an anomaly-free set with 
$n_g=0$.  Under the assumed breaking $SU(7)\to SU(5)\times SU(2)'\times U(1)_7$ we find that these representations, as well as the $SU(7)$ adjoint, $\bf{48}$,  will decompose 
as\cite{Slansky:1981yr,Yamatsu:2015npn} 
\begin{eqnarray}
&~&\bf{\bar 7} \to  \bf{(\bar5,1)_{-2}}+\bf{(1,2)}_5 \nonumber\\
&~&\bf{21} \to  \bf{(5,2)}_{-3}+\bf{(10,1)}_4+\bf{(1,1)}_{-10}\nonumber\\
&~&\bf{35} \to  \bf{(5,1)}_{-8}+\bf{(10,2)}_{-1}+\bf{(\bar{10},1)}_6\nonumber\\
&~&\bf{48} \to \bf{(24,1)}_0+\bf{(1,3)}_0+\bf{(1,1)}_0+\bf{(5,2)}_7+\bf{(\bar 5,2)}_{-7} \,.                
\end{eqnarray}
We observe that, as usual, the adjoint of Higgs in the $\bf{48}$ is responsible for both the initial $SU(7)$ breaking as well as that of the standard $SU(5)$ via the $\bf{(24,1)}_0$ component 
which we can imagine takes place at a comparable scale.  We also find as expected that the potential fermionic PM fields are necessarily {\it chiral} with respect to both $SU(2)'$ and $U(1)_7$ 
gauge groups. 

Here we see that it is easy to identify one linear combination of the $\bf{(\bar5,1)}_{-2}$ fields appearing in the $\bf{\bar 7}$'s with the usual $\bf{\bar 5}$ of $SU(5)$ that is an $SU(2)'$ singlet 
and thus automatically will have $Q_D=0$ since $Q_7$ does not contribute to this quantity as discussed above. 
In case $(b$), we need to identify the usual $\bf{10}$ containing SM fields with $\bf{(10,2)}_{-1}$ in the $\bf{35}$, 
but this representation is an $SU(2)'$ doublet, both of whose members must carry a value of $Q_D\sim \lambda_3 \neq 0$; this excludes the case ($b$) as a realistic possibility. In case $(a)$, 
the corresponding identification of the $\bf{(10,1)}_4$ in the $\bf{21}$ with the $\bf{10}$ 
of the usual $SU(5)$ avoids this particular issue since it automatically has $Q_D=0$ and these same types of choices would need to also be made elsewhere. 
We'll return to this issue below. 

The next, 
somewhat correlated, pair of obstacles we face are the generation of the various fermion masses as well as the breaking of $G_{Dark}$ (hopefully) down to $U(1)_D$. The required Higgs 
fields to do these two jobs can be found from among the same set of representations given above for the fermions but with the particular choices dependent upon which scenario 
$(a)$, $(c)$ or $(d)$ is being considered. Symbolically, in all three of these cases, such Higgs-Yukawa terms take the generic form (or some subset thereof) of the products of couplings 
\begin{equation}
\sim \bf {\bar {7}\cdot 35\cdot 21_H} +\bf {\bar {7} \cdot 21 \cdot \bar{7}_H}+\bf{21\cdot 35\cdot 21_H}+\bf{21\cdot 21 \cdot 35_H}+\bf{35\cdot 35\cdot 7_H}+\rm{h.c.}\,,
\end{equation}
where the subscript `H' labels a Higgs representation. As promised, here we see the first representative examples of {\it scalar} PM fields, carrying both dark and SM charges, as 
necessary ingredients to the overall gauge symmetry breaking and mass generation process. For all of these cases, however, one will generally be attempting to pair up (at least some of), \eg, 
the additional $\bf{(\bar5,1)}_{-2}$'s in the $\bf{\bar 7}$'s with the $\bf{(5,2)}_{-3}$'s in the $\bf{21}$ via a $\bf{(1,2)}_5$ from a Higgs in a $\bf{7_H}$ to generate vector-like mass terms for 
all of the PM fermions. Since these representations contain fields which are SM color-triplets, we know from previous 
analyses\cite{Rizzo:2018vlb,Rueter:2019wdf,Kim:2019oyh,Rueter:2020qhf,Wojcik:2020wgm,Rizzo:2021lob,Rizzo:2022qan,Wojcik:2022rtk,Rizzo:2022jti} recasting LHC searches that 
such states must have 
masses which are in excess of $\sim 1-2$ TeV so that the single $SU(2)'$-breaking vev of $\bf{7_H}$'s must be at least several TeV. Now this $SU(2)'$ doublet vev will also break both 
$SU(2)'$ as well as $U(1)_7$ but, as is well-known\cite{Li:1973mq}, will {\it not} leave any remaining unbroken subgroup of $G_{Dark}$ that can be identifiable as $U(1)_D$ since $U(1)_D$ must 
be a subgroup of $SU(2)_D$ by the discussion above.{\footnote{Remember that $Q_D$ does not have a contribution from the $U(1)_7$ charge as discussed above.}}. However, if two different 
$SU(2)'$ doublets obtain vevs which are not 'aligned' (\ie, both $T_3'=\pm 1/2$ members obtaining vevs even if this is in different doublets) in $SU(2)'$ space then even this remaining $U(1)$ will 
also be broken. Since both members of of the $\bf{(1,2)}_5$ $SU(2)'$ doublet necessarily carry $Q_D\neq 0$ even this single vev will break $U(1)_D$ at a high scale, violating our 
requirements above.

These initial considerations will then exclude case $(a)$ 
immediately without any further analysis but some more straightforward checking is required to see what happens in the cases $(c)$ and $(d)$. It doesn't take 
long, however, to convince oneself that in both of these cases a $\bf{(1,2)}_5$ from a Higgs in a $\bf{7_H}$ will still be required to give masses to some set of non-SM, vector-like fermions 
at or above the TeV mass scale. For example, the $\bf{(10,2)}_{-1}+\bf{(\bar{10},1)}_6$ representations in the $\bf{35}$ will pair up to form such vector-like states via the vev of the 
$\bf{(1,2)}_{-5}$,  $SU(2)'$ iso-doublet in the $\bf{7_H}$ in both scenarios $(c)$ and $(d)$. Since these fields also contain VL-color-triplet fermions that need large masses, by our previous 
discussion this excludes these cases as well since this vev necessarily breaks $U(1)_D$. Extending these arguments to their logical conclusion we find that the identification of $G=SU(7)$ 
is a failure, since $SU(2)'$ doublet vevs are always required, leaving $U(1)_D$ broken at a large scale. 
It is interesting (and unfortunate) to note that if we had {\it not} needed these $SU(2)'$ doublets to generate the PM vector-like fermion masses we could have simply employed the 
$\bf{(1,3)}_0$ in the $\bf{48}$ to break the $SU(2)'$ gauge symmetry and this would have left us with an unbroken $U(1)_D$ having the desired charge assignments.

\subsection{$SU(8)$}

As in the case of $SU(7)$, $G=SU(8)$ offers many model building opportunities that have been discussed from time to time over the last few decades but which we can still divide into 
$n_g=1$ and $n_g=3$ subsets. Note that in this scenario, we recall that $Q_D=a_1\lambda_3+a_2\lambda_8$ since now $G_{Dark}=SU(3)'\times U(1)_8$ with $SU(3)'$ being rank 2, 
allowing for the possibility of a single $Q_D=0$ field within an $SU(3)'$ $\bf{3/\bar 3}$ representation. Here, we'll see another example of how things can go wrong which will also be common 
for the larger unification groups we encounter below.

Some of the most common, but yet not exhaustive set of example $SU(8)$ models appearing in the literature,
\eg, \cite{Georgi:1979md,Frampton:1979cw,Vaughn:1979sm,Kim:1980qi,Kim:1981vy,Deshpande:1980fx,Chkareuli:1992kd,Barr:2008pn} 
are given in the following list:
 \begin{eqnarray}
 (a)&~~~~& 3[4(\bf {\bar 8})+\bf {28}]\nonumber\\
 (b)&~~~~& 3[\bf{\bar 8}+\bf{\bar{28}}+\bf{56}]\nonumber\\
 (c)&~~~~& 3[2(\bf{8})+3(\bf {\bar {28}})+2(\bf{56})]\nonumber\\
 (d)&~~~~& 5(\bf{\bar {28}})+4(\bf{56})\nonumber\\
 (e)&~~~~& 9(\bf{\bar 8}) +\bf {28}+\bf {56}\,,
 \end{eqnarray} 
where the first three obviously have $n_g=1$ while the last two have $n_g=3$. As before, other possibilities can be obtained by adding/subtracting `multiples' of the combination of 
representations $3\bf{(\bar 8})+2(\bf{28)}+\bf{\bar{56}}$\cite{Fonseca:2015aoa} which itself forms an anomaly-free set with $n_g=0$. Under the $SU(8)\to SU(5)\times SU(3)'\times U(1)_8$ 
decomposition one finds for the relevant representations that 
\begin{eqnarray}
&~&\bf{\bar 8} \to \bf{(\bar 5,1)_{-3}}+\bf{(1,\bar 3)}_5\nonumber\\
&~&\bf{28} \to  \bf{(5,3)_{-2}}+\bf{(10,1)}_6+\bf{(1,\bar 3)}_{-10}\nonumber\\
&~&\bf{56} \to  \bf{(5,\bar 3)}_{-7}+\bf{(10,3)}_{1}+\bf{(\bar{10},1)}_9+\bf(1,1)_{-15}\nonumber\\
&~&\bf{70} \to  \bf{(5,1)}_{-12}+\bf{(10,\bar 3)}_{-4}+\bf{(\bar{10},3)}_4+\bf(\bar 5,1)_{12}\nonumber\\
&~&\bf{63} \to \bf{(24,1)}_0+\bf{(1,8)}_0+\bf{(1,1)}_0+\bf{(5,\bar 3)}_8+\bf{(\bar 5,3)}_{-8} \,.                
\end{eqnarray}
Note that the $\bf{70}$ is a real representation as, of course, is the $\bf{63}$ adjoint. For $N=8$, as noted above, the asymptotic freedom requirement now starts to be felt in a non-trivial 
way since several of these possibilities might fail immediately without even considering the potential scalar contributions to 
the $\beta$-function which will only make matters worse: we find, however, that only case ($c$) fails this requirement when we take 3 multiples of a single set of representations to obtain the 
three SM generations and so it no longer needs to be realistically considered in the discussions that follow. 

The gauge symmetry breaking in the $SU(8)$ scenario is rather familiar with the $SU(5)$ $\bf {24}$ performing its usual role. A single fundamental $\bf{3/\bar 3}$ will break $SU(3)'$ down 
to $SU(2)'$\cite{Li:1973mq} while the vev of the $T_3'=0$ member of the $SU(2)'$ triplet within the adjoint, $\bf{(1,8)}_0$, will then break $SU(2)'$ down to $U(1)_D$ as desired. We note, 
however, that multiple fundamental $\bf{3,\bar 3}$'s whose vevs are not `aligned', \ie, not all having $Q_D=0$,  
will result in $SU(3)'$ breaking completely {\it without} any surviving $U(1)$'s\cite{Li:1973mq} below the electroweak/TeV scales that we can identify with $U(1)_D$ as is required by our 
model building constraints above. Note that {\it only} scalar fields in the $\bf{3,\bar 3}$ representations appear that can break $SU(3)'$ here. 
Further, since several distinct (yet vev aligned) fundamentals appear with different values of the 
$U(1)_8$ charge, this symmetry will break at the same mass scale as does $SU(3)'$; this will be a common feature for all the models below with $G=SU(N), N\geq 8$. We again note that the 
PM fields are chiral under the unbroken $SU(2)'$ group. 

Turning to the fermion mass terms, as we might expect, the required Higgs fields (apart from the usual adjoint) are essentially also members of this same set of representations as 
are the fermions. Similarly to the $SU(7)$ case above, we can again symbolically write the Higgs-Yukawa interaction terms for the fermion masses in a generic form (or as some subset 
thereof depending upon the case) of the products
\begin{equation}
\sim \bf {\bar {8}\cdot 28\cdot \bar{8}_H} +\bf {\bar {8}\cdot 56 \cdot \bar{28}_H}+\bf{28\cdot 56\cdot 56_H}+\bf{56\cdot 56 \cdot 28_H}+\bf{28\cdot 28\cdot 70_H}+\rm{permutations}+\rm{h.c.}\,,
\end{equation}
so that the number of possible PM and SM fermion mass generation terms are each somewhat restricted in all cases. 

Note that in case ($a$), we can select one linear combination the four $\bf{(\bar5,1)_{-3}}$ fields contained in the four $\bf{\bar{8}}$'s to be the `conventional' SM $\bf{\bar 5}$ of $SU(5)$ 
while also choosing the $\bf{(10,1)}_6$ from the $\bf{28}$ as the usual $\bf{10}$. Since both of these fields are already $SU(3)'$ singlets, the first issue we had to deal with in the $SU(7)$ 
model above is trivially bypassed and the SM fields will have $Q_D=0$ automatically as required. Simultaneously, the Higgs fields needed to supply 
vector-like masses to the non-SM fermions are now either in singlets or in $\bf{3}$/$\bar{\bf{3}}$'s of $SU(3)'$ as we can tell from the representation decompositions above. 

Case ($a$) has a somewhat simple symmetry breaking sector as, apart from the adjoint $\bf {63}$, it only requires the first and fifth terms in Eq.(10), \ie, Higgs fields in both the 
$\bf {\bar {8}_H}$ and 
$\bf{70_H}$ representations and, since the three families are a simple replication of one subset, we can consider for simplicity fermions in a single combination of $4(\bf{\bar 8})+\bf {28}$ fields. 
Thinking at the $SU(5)$ level we can identify one linear combination of the four $\bf {(\bar 5,1)_{-3}}$'s as the `SM field' while the other three $\bf{(\bar 5,1)_{-3}}$'s must then match up with the 
$\bf{(5,3)_{-2}}$ to form the vector-like PM fields. From this we learn two things: ($i$) the SM fermion masses are necessarily generated by two distinct SM $SU(2)_L$ Higgs 
isodoublets, one from the $\bf {\bar 8_H}$ and the other from the $\bf {70_H}$ with the same type of coupling structure that occurs in the type-II Two Higgs Doublet 
Model\cite{Gunion:1989we}. ($ii$) Even 
more importantly, the PM mass term in this model must necessarily be of the general form (in obvious $53'1_8$ language): 
\begin{equation}
y_{ia}~[\bf {(\bar 5,1)_{-3}}]_i~\bf{(5,3)_{-2}}~[\bf{(1,\bar 3)}_5]^a_H +\rm{h.c.}\,,
\end{equation}
where the $y$'s are Yukawa couplings, the index $i=1-3$ labels the three remaining $\bf {\bar 5}$'s and here we will allow for the possibility of more than one relevant Higgs anti-triplet, labelled 
by the index $a=1,...$. 

First consider the simplest case when only a single Higgs field ($a=1$) is present;  since only one element of the $[\bf{(1,\bar 3)}_5]_H$ is allowed to have a ($Q_D=0$) vev, this projects 
out a single corresponding element in the $\bf{(5,3)_{-2}}$ implying only a single fermion bilinear can be constructed in the $SU(3)'$ subspace so that in the full $53'1_8$ space only a 
single set of five fermion bilinears can obtain masses, \ie, one of the $\bf {\bar 5}\cdot \bf{5}$'s obtains a mass term and only five (degenerate) fermion masses are the result. If 
we increase the number of Higgs fields and allow for {\it arbitrary} alignment of their vevs, then three Higgs fields will generate all the desired mass terms.  However, as is well-known, 
in such a situation the $SU(3)'$ group breaks completely\cite{Li:1973mq} leaving us without a low-energy $U(1)_D$ gauge group. If instead, we add extra Higgs fields  
where all the vevs occurs in the same element of the representation so that they are aligned, as is required to that $U(1)_D$ remains unbroken, this will not alter the result obtained with only a 
single Higgs field with only one $\bf {\bar 5}\cdot \bf{5}$ mass term resulting. Thus, it is impossible to generate tree-level 
masses for these remaining two candidate PM fields at the $SU(3)'$-breaking scale. 
This is a disaster for this case because at the subsequently $SU(2)'$-breaking scale (which must lie above a few TeV), to avoid breaking $U(1)_D$ while also generating the required 
gauge boson masses, \ie, those apart from that of the dark photon, the $T_3'=0$ member of the real $SU(2)'$ triplet in in the adjoint is employed. Giving a $T_3'\neq 0$ member of any 
scalar representation a vev will automatically break $U(1)_D$ at/above the few TeV scale and, thus, at least at tree-level, the remaining $\bf {\bar 5}\cdot \bf{5}$ terms must be absent. 
Since not all the needed masses can be generated at the $SU(3)'$ breaking scale, case ($a$) is excluded; this will be a very common feature of the many scenarios we will encounter below. 

In case ($e$), since $n_g=3$, three linear combinations of the nine $\bf {\bar 8}$'s contain the three $\bf {\bar 5}$'s which are also $SU(3)'$ singlets that we must identify as 
$Q_D=0$ SM fields while the 
$\bf 28$ contains a $\bf (10,1)_6$ which we can also identify with a SM $Q_D=0$ field. However, we still need two more $SU(5)$ $\bf 10$'s with $Q_D=0$ to identify with the remaining SM 
fields and the only possible source for these lies within the $\bf(10,3)_1$ in the $\bf 56$ which is an $SU(3)'$ triplet. This requires that when the operator $Q_D=a_1\lambda_3+a_2\lambda_8$ 
acts on this triplet it produces two zero eigenvalues by a suitable choice of $a_{1,2}$. But, of course as we know, this can't happen as at most one zero eigenvalue can be obtained. This 
implies that one of the SM generations 
necessarily carries $Q_D\neq 0$ which is not phenomenologically acceptable and violates our model-building assumptions above, thus excluding case ($e$). 

The situation is found to be quite different in case ($d$) where we see immediately that {\it none} of the usual $SU(5)$ $\bf{\bar 5}$'s or $\bf{10}$'s are $SU(3)'$ singlets. However,  we are able to 
freely choose one component of these triplet/anti-triplet representations to have $Q_D=0$; we can, without loss of generality, further take this to always be, \eg,  
the lowermost component in such triplet fields 
for purposes of this discussion. Then we see that while a consistent pair of mass terms may be obtainable for the choice of SM fields, we cannot simultaneously generate $\sim$ few TeV-scale 
$SU(3)'$-breaking masses required for {\it all} of the potential vector-like PM fermions in the remaining set of $\bf{ 5+\bar 5}$ and $\bf {10 +\bar {10}}$ representations without also breaking the 
$U(1)_D$ gauge symmetry, as was seen in case ($a$), as the required Yukawa mass terms will take the symbolic form
\begin{equation}
\sim \bf{(10,3)_1\cdot (\bar{10},1)_9\cdot (1,\bar 3)^H_{-10}}+\bf{(\bar 5,\bar 3)_2\cdot (5,\bar 3)_{-7}\cdot (1,\bar 3)^H_5}+\rm{h.c.}\,.
\end{equation}
Although there are two different species of anti-triplet Higgs fields appearing here, their vevs must still be aligned along the $Q_D=0$ in direction as we recall that only a single component of any 
of  the triplet/anti-triplet Higgs fields can obtain a vev if we want to obtain an unbroken $U(1)_D$. This apparently excludes case ($d$). 

However, maybe we can obtain some additional freedom in this particular case by recalling the caveat we noted above about the requirement that $Q_D$ can only be some linear combination 
of the diagonal $SU(N-5)'$ generators, omitting any possible contribution for $U(1)_N$, \ie, above being just $Q_D=a_1\lambda_3+a_2\lambda_8$, but now allowing for an additional 
$Q_8$ contribution.  Perhaps we can 
apply this to case ($d$) as {\it neither} of the fields that we identify with the SM $(\bf {\bar 5 +10})$ are $SU(3)'$ singlets but are instead triplets/anti-triplets,
\ie, by taking the SM fermions to be just the $Q_D=0$ members of three copies of $ \bf{(\bar 5,\bar 3)_{2}} +\bf{(10,3)}_{1}$. 
But then we immediately see that the Higgs responsible for generating the masses  
of the $d,e$-type SM fermions must be the $SU(3)'$ singlet field $\bf{(\bar 5,1)_{-3}}$ which is required to have $Q_D=0$ so that $U(1)_D$ survives unbroken below the electroweak scale - 
yet this Higgs scalar carries a non-zero value of $Q_8=-3$. Thus it remains true that $Q_D$ cannot have a contribution from the $U(1)_8$ generator, $Q_8$, even in this case. 
Furthermore, one finds that even if we allow for the possibility that the $Q_D=0$ element is in different locations within the (anti-)fundamental representation depending upon the value of 
$Q_8$, this result persists. At the very least 
we see that, in general, when $G=SU(N)$, if a Higgs representation which is needed to generate SM particle masses is an $SU(5-N)'$ singlet it is not allowed to carry a non-zero value of 
$Q_N$ if $Q_N$ contributes to $Q_D$. It is interesting to note that in the already excluded case ($c$), something very similar happens where the $\bf{(\bar 5,1)^H_{-3}}$ 
must generate the SM $d,e$-type mass terms so that $Q_D$ must again be independent of $Q_N$. Further, we se that a further necessary, but not sufficient, condition to allow for a 
$Q_N$ contribution to $Q_D$ is to make sure that the $\bf {\bar 5}$ and $\bf{10}$ SM fields do not transform as conjugate representations under the $SU(N-5)'$ group so that the 
Higgs field(s) needed to generate mass terms are not $SU((N-5)'$ singlets. Similar arguments can be applied to the other fields as well.

Now having ruled out this case, as well as ($e$) above, we see that all of the $n_g=3$ scenarios are excluded. It is also clear from this discussion that the survival threshold is set higher for 
these model varieties than for those with $n_g=1$; this will continue to be the case as we increase $N$ as will be seen below.

In the remaining case ($b$), the symmetry breaking requirements can be somewhat more complex since, while the usual SM $SU(5)$ $\bf {10}$ must be identified with the lower member of 
the $\bf{(10,3)_1}$ in the $SU(8)$ 
$\bf{\bar{28}}$, the SM $\bf{\bar 5}$ can {\it either} be the $\bf{(\bar 5,1)_{-3}}$ in the $\bf{\bar 8}$, as in case ($a$), or the lower member of the $\bf{(\bar 5,\bar 3)_2}$ in the $\bf{\bar {28}}$.  
Both of these assignments are made possible by assuming, \eg, that the lower members of all triplets/anti-triplets have $Q_D=0$ as we did in case ($d$). Both of these choices allow for 
the generation of the SM 
$d$-quark and charged lepton masses via either $\bf{(\bar 5,1)_{-3}\cdot (10,3)_1\cdot (\bar 5,\bar 3)^H_2}$-type or $\bf{(\bar 5,\bar 3)_2\cdot (10,3)_1\cdot (\bar 5,1)^H_{-3}}$-type 
Yukawa couplings. Note that in this later case the fields in the $\bf{(\bar 5,\bar 3)_2}$ with $Q_D\neq 0$ can also pick up an electroweak-scale mass term. Recall that in either of these cases, the SM fermion or Higgs representation assignments will still force us to require that $Q_D=a_1\lambda_3+a_2\lambda_8$ without there being any $Q_8$ contribution. For either choice of these 
assignments, the $u$-quark mass is always generated by the coupling $\bf{(10,3)_1 \cdot (10,3)_1 \cdot (5,3)^H_2}$, where again both $Q_D=0$ as well as the $Q_D\neq 0$ components 
can pick up electroweak-scale masses. 

The most significant, yet as we have seen apparently common, problem one faces with case $(b)$ is the lack of a sufficient number of mass terms for all of the PM fields which should lie in 
(at least) the few TeV range. The only mass terms for pairs of $SU(5)$ non-singlets fermions that are generated by $SU(5)$-singlet, $SU(3)'$-breaking (\ie, non-singlet) Higgs scalars 
with potentially large vevs are seen to be of the general coupling structures: 
\begin{equation}
\sim \bf{(10,3)_1\cdot (\bar{10},1)_9\cdot (1,\bar 3)^H_{-10}}+\bf{(\bar 5,1)_{-3}\cdot (5,\bar 3)_{-7}\cdot (1,3)^H_{10}} +\bf{(\bar 5, \bar 3)_2\cdot (5,\bar 3)_{-7}\cdot (1,\bar 3)^H_5}
+\rm{h.c.}\,.
\end{equation}
Employing this expression it is easily seen that it is impossible to simultaneously supply all of the desired PM mass terms while also keeping $U(1)_D$ unbroken below the few TeV scale since 
all the $\bf{3,\bar 3}$ vevs must all be aligned in a single direction, so that 
this case is also excluded. 

From the set of analyses above we can conclude that all of the $SU(8)$ scenarios that we have considered are excluded thus disfavoring this unification gauge group.

\subsection{$SU(9)$}

Since in the case of $G=SU(9)$ one has $G_{Dark}=SU(4)'\times U(1)_9$, here we can define the dark charge as $Q_D=a_1\lambda_3+a_2\lambda_8+a_3\lambda_{15}$, so that a 
$\bf{4,\bar 4}$ representation can now have up to two of its members with $Q_D=0$ thus generalizing the case of $SU(3)'$ seen above. These two potential $Q_D=0$ elements can be easily 
achieved by, \eg, taking 
$a_2=a_3=0$ although this is not a unique choice.  Thus, to break the $SU(4)'$ gauge symmetry in this case we can in principle employ two un-aligned vevs in different $\bf{4}$ or $\bf{\bar 4}$'s 
to reduce the symmetry to $SU(2)'$ and then follow the same path as was discussed above for the $SU(8)$ scenario employing the $SU(2)'$ real triplet in the $SU(9)$ adjoint to reduce this 
gauge symmetry further down to $U(1)_D$.  Note also that, unlike the previously considered models, the Higgs fields in the (now distinct) second rank, anti-symmetric tensor representation 
(or its conjugate) of $SU(4)'$ , \ie, $\bf{6,\bar 6}$, can also participate in the symmetry breaking process, which can break $SU(4)'$ directly down to $SU(2)'$\cite{Li:1973mq}. We see that 
the chosen number of $Q_D=0$ elements of the fundamental representation 
will then determine the corresponding number of such elements in the $\bf{6,\bar 6}$ second rank anti-symmetric representation, \ie, for a single $Q_D=0$ vev to occur for such reps 
here we need to have two $Q_D=0$ elements in the fundamental. As 
in the case of $SU(8)$, $U(1)_9$ will break at the same scale as does $SU(4)'$ since several SM singlet dark Higgs fields will always be present with different values of $Q_9$.

As in the examples of both $SU(7)$ and $SU(8)$ above, there are many representative $SU(9)$ unification models that have been considered in the previous 
literature\cite{Georgi:1979md,Frampton:1979cw,Vaughn:1979sm,Fujimoto:1981iu,Kang:1985ms,Frampton:2009ce,Dent:2009pd} having either $n_g=1$ or 3, \eg,
 \begin{eqnarray}
 (a)&~~~~& 3[5(\bf {\bar 9})+\bf {36}]\nonumber\\
 (b)&~~~~& 3[\bf{\bar {36}}+\bf{126}]\nonumber\\
 (c)&~~~~& 3[\bf{\bar 9}+\bf{\bar {84}}+2(\bf{126})]\nonumber\\
 (d)&~~~~& 4(\bf{\bar 9})+2(\bf {36})+\bf{84}+\bf{\bar {126}}\nonumber\\
 (e)&~~~~& 9(\bf{\bar 9}) +\bf {84}\,.
 \end{eqnarray} 
which form a non-exhaustive but typical set of these possibilities. As before, this set is easily expandable by, \eg,  `adding/subtracting' multiples of the 
anomaly-free combinations\cite{Fonseca:2015aoa} 
$6(\bf{\bar 9}) +3(\bf{36})+\bf{\bar {84}}$ and/or $5(\bf{\bar 9}) +2(\bf{36}) +\bf{\bar {126}}$. However, when doing so one must take care that the asymptotic freedom requirement imposed 
on $SU(9)$ is still satisfied since, as noted previously,  the strength of this requirement grows as $N$ increases.  Further, under the $SU(9)\to SU(5)\times SU(4)'\times U(1)_9$ 
decomposition, one finds for the relevant representations that 
\begin{eqnarray}
&~&\bf{\bar 9} \to \bf{(\bar5,1)_{-4}}+\bf{(1,\bar 4)}_5\nonumber\\
&~&\bf{36} \to  \bf{(5,4)}_{-1}+\bf{(10,1)}_8+\bf{(1,6)}_{-10}\nonumber\\
&~&\bf{84} \to  \bf{(5,6)}_{-6}+\bf{(10,4)}_{3}+\bf{(\bar{10},1)}_{12}+\bf(1,\bar 4)_{-15}\nonumber\\
&~&\bf{126} \to  \bf{(5,\bar 4)}_{-11}+\bf{(10,6)}_{-2}+\bf{(\bar{10},4)}_7+\bf(\bar 5,1)_{16}+\bf{(1,1})_{-20}\nonumber\\
&~&\bf{80} \to \bf{(24,1)}_0+\bf{(1,15)}_0+\bf{(1,1)}_0+\bf{(5,\bar 4)}_9+\bf{(\bar 5,4)}_{-9} \,.                
\end{eqnarray}
As was also found with the $G=SU(8)$ possibility, we see that some of these models above will immediately fail the growingly powerful asymptotic freedom condition for $SU(9)$ - even without 
any consideration of the scalar sector. Here we observe that we no longer need to realistically consider cases ($b$) and ($c$) any further, which is rather unfortunate, as the set of 
three copies of the single set of representations 
needed to reproduce the three SM generations has just too many degrees of freedom to maintain AF above the unification scale - even before the symmetry breaking Higgs scalar sector 
contributions are included in the $SU(9)$ $\beta$-function. Case ($d$) may also be excluded by this constraint depending upon the required complexity of its scalar Higgs sector as we 
shall soon find. We do note in passing that in case ($c$), even though the SM fields are embedded in a non-trivial fashion, the Higgs field responsible for generating the $u$-type quark masses 
is an $SU(4)'$ singlet carrying $Q_9\neq 0$ thus disallowing any contribution of $Q_9$ to $Q_D$. 

To go further, we note that the various mass terms that may be needed in this $SU(9)$ scenario will be generated by (some subset of) the general Yukawa terms resulting from the 
generic combination 
\begin{equation}
\sim \bf {\bar {9}\cdot 36\cdot \bar{9}_H}+\bf{{\bar 9}\cdot 126\cdot \bar{84}_H}+ 
\bf {\bar {9}\cdot 84\cdot \bar {36}_H} +\bf{36 \cdot 36 \cdot \bar{126}_H}+ \bf{36\cdot 84 \cdot 126_H} + \bf{84\cdot 84\cdot 84_H}+...+\rm{perms.}+\rm{h.c.}\,.
\end{equation}
In case ($d$), since the fermion fields themselves lie in representations of four different dimensionalities, the Yukawa interaction expression above tell us that we'll need at least one Higgs scalar 
in each of the $\bf{9,36,84}$ and $\bf{126}$ (or their conjugates) representations as well as the adjoint $\bf{80}$ to break all the symmetries. This increases the $SU(9)$ beta function by a 
positive factor of $\delta \beta= (9+n_9+7n_{36}+21n_{84}+35n_{126})/6$ so that even if only one of each type of Higgs scalar representation were introduced the additional degrees of 
freedom would also render this case non-asymptotically free and so it too now becomes excluded.

For case ($a$), only a few of these possible mass generating terms above are relevant. 
From this subset we see that case ($a$) for $SU(9)$ is indeed quite similar to case ($a$) for $SU(8)$ where (for a single generation) we identify one linear combination of the 
five $\bf(\bar 5,1)_{-4}$ fields from the five $\bf{\bar 9}$'s with the usual SM field yet the assignment of the $\bf(10,1)_8$ from the $SU(9)$ $\bf {36}$ is unique. While there are then 
no issues with the SM fermion masses, since these fields are themselves $SU(4)'$ singlets as above, problems do arise in generating the masses for the PM fields which transform 
non-trivially under both $SU(5)$ and $SU(4)'$, similar to what was found in the case of $SU(8)$. One finds that the relevant mass generating couplings in the present case to be of the form  
\begin{equation}
y_{ia}~[\bf {(\bar 5,1)_{-4}}]_i~\bf{(5,4)_{-1}}~[\bf{(1,\bar 4)}_5]^a_H +\rm{h.c.}\,,
\end{equation}
where the $y$'s are Yukawa couplings as above while the index $i=1-4$ now labels the four remaining combinations of the $\bf {\bar 5}$'s and we again allow for the possibility of more than 
one Higgs field, labelled by the index $a$. Since only the two $Q_D=0$ components of any $SU(4)$ $\bf 4$ or $\bf {\bar 4}$ fields can obtain vevs without also breaking $U(1)_D$, two 
of the desired fermionic PM fields will not be able to obtain TeV-scale masses this way.  Thus case ($a$) is found to also be 
excluded when $G=SU(9)$ as was expected from the same arguments made previously for the case ($a$) of $SU(8)$. 

Lastly, we must consider the interesting case ($e$) which benefits from having fermion fields in only two distinct types of representations but whose symmetry breaking requirements are 
rather complex since it is an $n_g=3$ scenario with both SM and candidate PM fields lying in the same representations at the $SU(5)\times SU(4)'\times U(1)_9$, \ie, $54'1_9$ level.  As can 
be seen from the expression above this implies that only the $\bf{36_H,84_H}$-type Higgs scalars (and/or their complex conjugates) are needed to generate all the fermion mass terms which 
helps to satisfy the asymptotic freedom constraint.  However, this also implies that the fermion fields obtaining their masses in the breaking of $SU(4)'$ might be `mixed' in a non-trivial manner 
with those receiving masses due to the usual SM electroweak symmetry breaking. Fortunately, we can remove some of this complexity by working in the approximate 
limit where the $SU(4)'$ breaking scale is taken to be much larger than the electroweak scale so that these two symmetry breaking steps can be effectively decoupled.
On the other hand, making matters more complicated are the two types of representations of pure dark sector sets of fields,  
$\bf(1,\bar 4)_{5,-15}$, which also enter into the fermion mass matrix at the same $SU(4)'$-breaking mass scale.  For example, while three sets of the $\bf(\bar 5,1)_{-4}$'s from the nine 
$\bf{\bar 9}$'s might be identified with usual $Q_D=0$, $SU(5)$ SM $\bf \bar 5$ fields, the remaining six of these must then pair up with the $\bf{(5,6)_{-6}}$ in the $\bf{84}$ via 
a $\bf [(1,6)_{10}]_H$ 
in the $\bf{84_H}$ to form massive PM fields.  Similarly, three of the members of the $SU(4)'$ $\bf 4$ representation in the $\bf (10,4)_3$'s must be identified with usual $SU(5)$ SM $\bf 10$ 
fields while the remaining multiplet member must pair up with the corresponding $\bf{(\bar {10},1)_{12}}$ via a $\bf[(1,\bar 4)_{-15}]_H$, also in the $\bf{84_H}$, to form additional PM.  Once 
the Higgs scalar in the coupling $\bf{(10,4)_3\cdot (\bar {10},1)_{12}\cdot [(1,\bar 4)_{-15}]_H}$ obtains a (single) vev, breaking $SU(4)'$ down to $SU(3)'$, one sees that one of these $\bf 10$'s 
indeed obtains a large mass above the TeV scale as is needed. While this last step was rather straightforward, the rest of the required mass generation is found to be somewhat more 
difficult to manage given our model building constraints. 

As we have seen above, the requirement that only the $Q_D=0$ elements of the various Higgs representations are allowed to obtain vevs (at the electroweak scale or above) so that 
$U(1)_D$ remains unbroken leads to the result that many fields remain massless at the required breaking scale. As a practical example in the present case, consider the mass term for the 
six sets of the $\bf(\bar 5,1)_{-4}$'s mentioned above that we want to obtain masses at the $SU(4)'$ breaking scale by pairing with up those in the the single $\bf{(5,6)_{-6}}$ via a 
Yukawa coupling structure similar to what we have frequently encountered above, \ie, 
\begin{equation}
y_{ia}~[\bf {(\bar 5,1)_{-4}}]_i~\bf{(5,6)_{-6}}~[\bf{(1,6)}_{10}]^a_H +\rm{h.c.}\,,
\end{equation}
where $i=1-6$ now runs over these six non-SM $\bf{\bar 5}$'s as before. While the matching numbers of degrees of freedom are correct for generating mass terms for all of these states, the vev 
structure of the single $\bf [(1,6)_{10}]_H$ (which has only two $Q_D=0$ elements) is insufficient to accomplish this, allowing for at most two of these pairings to pick up masses as the  
multi-TeV level. Note that adding additional $\bf [(1,6)_{10}]_H$'s won't be helpful as long as only the $Q_D=0$ elements are allowed to obtain vevs so this structure is necessarily aligned. 
Further, we note that at the level of the multi-TeV or above $SU(4)'$ breaking scale, the additional dark sector fields in the two $\bf (1,\bar 4)$ do not enter into these considerations. Similarly, 
we face essentially the same difficulties when we want to give $SU(2)_L$-breaking masses to the three remaining $\bf{10}$'s discussed above.  Here we are faced with the Yukawa couplings 
\begin{equation}
\sim \bf{(10,4)_3}~\bf{(10,4)_{3}}~[\bf{(5,6)_{-6}}]_H +\rm{h.c.}\,,
\end{equation}
which is again seen to be insufficient to able to perform the required role, generating only two (instead of the needed three) mass terms for the up-like quarks. Thus, apparently, case 
$(e$) is also excluded from our list of candidate scenarios due to the now common `insufficient number of vevs' problem.  

We can conclude from this discussion above that the choice $G=SU(9)$ is highly disfavored or more likely excluded.

Before continuing we make the following observation: 
interestingly, we find that as $N$ increases one sees that that the number of $Q_D=0$ vevs that are allowed in the (anti-)fundamental and higher scalar representations also increases. 
However, we also see that, apparently, it does not increase fast enough (at least in the present example) to generate all the required masses for the SM fields as well as for the similarly 
growing number of candidate PM fermions that we encounter.

\subsection{$SU(10)$}

In the scenario with $G=SU(10)$ broken by the vev of the Higgs in the adjoint representation, one has $G_{Dark}=SU(5)'\times U(1)_{10}$, and thus we can define the dark charge 
as $Q_D=a_1\lambda_3+a_2\lambda_8+a_3\lambda_{15}+a_4\lambda_{24}$ This implies  
that a $\bf{5/\bar 5}$ representation can now have up to 3 members with $Q_D=0$ which may (or may not) be fortuitously the same as the number of SM generations. As in previous cases, 
since $SU(5)'$ singlet fields carrying non-zero values of the $U(1)_{10}$ charge, $Q_{10}$, are involved in generating the SM fermion masses, under the assumption that $Q_D(SM)=0$, then 
$Q_D$ cannot have a contribution from $Q_{10}$. Similar to the discussion above, $Q_D=0$ for three members of the fundamental can be easily achieved by, \eg, taking $a_2=a_3=a_4=0$ 
although this is far from a unique choice.  To break the $SU(5)'$ gauge symmetry in this case, we can in principle 
employ three distinct un-aligned vevs in different $\bf{5}$ or $\bf{\bar5}$'s reducing the symmetry to $SU(2)'$ as above, then employ the  $SU(2)'$ triplet in the $SU(10)$ adjoint to reduce  
this even further to the desired $U(1)_D$.  Here with three $Q_D=0$ members of the fundamental/anti-fundamental obtaining vevs (and hence breaking the $SU(3)'$ subgroup of the $SU(5)'$), 
the second-rank anti-symmetric $\bf{10/\bar {10}}$ Higgs representation will then correspondingly now have {\it four} vevs with $Q_D=0$, three of which contribute to $SU(3)'$ breaking while the 
fourth breaks any additional remaining $U(1)'$ so that only an $SU(2)'$ remains. As was the cases above, $U(1)_{10}$ will break at the same scale as does $SU(5)'$ since we always 
have several SM singlet dark Higgs fields 
present with different values of $Q_{10}$. The $SU(10)$ model is also unique amongst those previously considered as both the visible and dark sectors are (obviously) described by the same 
non-abelian gauge group. As in the models above, we note that the PM fields will be {\it chiral} with respect to $SU(5)'$ and, hence with respect to the $SU(2)'$ group just as the SM fields are 
chiral with respect to $SU(2)_L$. 

As in the cases above, many representative $SU(10)$ unification models that have been considered in the previous 
literature\cite{Georgi:1979md,Frampton:1979cw,Frampton:1979fd,Frampton:1979tj,Vaughn:1979sm,Langacker:1980js,Kashibayashi:1982nu} with either $n_g=1$ or 3, \eg,
 \begin{eqnarray}
 (a)&~~~~& 3[6(\bf {\bar {10}})+\bf {45}]\nonumber\\
 (b)&~~~~& 3[\bf{\bar {120}}+\bf{210}]\nonumber\\
 (c)&~~~~& 2(\bf {10})+5(\bf{\bar{45}})+2(\bf {120})\nonumber\\
 (d)&~~~~& 8(\bf{\bar {10}})+\bf {\bar {45}}+\bf{120}\nonumber\\
 (e)&~~~~& 2(\bf{\bar {10}}) +2(\bf{\bar {45}})+\bf{210}\,,
 \end{eqnarray} 
which form a non-exhaustive but representative set of these possibilities. As in the previous scenarios, this set is easily expandable by, \eg,  `adding/subtracting' multiples of the 
combinations\cite{Fonseca:2015aoa} $10(\bf{\bar {10}}) +4(\bf{45})+\bf{\bar{120}}$ and/or $16(\bf{\bar {10}}) +5(\bf{45}) +\bf{\bar {210}}$. However, in using this freedom one again 
quickly runs into difficulties  
with $SU(10)$'s asymptotic freedom as the representations are growing quite large - especially so if multiple $\bf{120/\bar{120}}$'s are needed or if more than a single $\bf{210/\bar{210}}$ is 
present as in case $(b$). Here, one must seriously take (in this case extra) care to insure that the asymptotic freedom requirement imposed on the $SU(10)$ gauge coupling is still 
satisfied, especially now with the noted large dimensionalities of these representations.  

We note that under the $SU(10)\to SU(5)\times SU(5)'\times U(1)_{10}$ decomposition, one finds for the relevant representations that 
\begin{eqnarray}
&~&\bf{\bar {10}} \to \bf{(\bar5,1)_{-1}}+\bf{(1,\bar 5)}_1\nonumber\\
&~&\bf{45} \to  \bf{(5,5)}_0+\bf{(10,1)}_2+\bf{(1,10)}_{-2}\nonumber\\
&~&\bf{120} \to  \bf{(5,10)}_{-1}+\bf{(10,5)}_{1}+\bf{(\bar{10},1)}_{3}+\bf(1,\bar {10})_{-3}\nonumber\\
&~&\bf{210} \to  \bf{(5,\bar{10})}_{-2}+\bf{(10,10)}_{0}+\bf{(\bar{10},5)}_2+\bf{(\bar 5,1)}_{4}+\bf{(1,\bar 5)}_{-4}\nonumber\\
&~&\bf{99} \to \bf{(24,1)}_0+\bf{(1,24)}_0+\bf{(1,1)}_0+\bf{(5,\bar 5)}_2+\bf{(\bar 5,5)}_{-2}\,,                
\end{eqnarray}
again showing the obvious symmetry between the visible and dark sectors.  As was found with both the $G=SU(8)$ and $G=SU(9)$ possibilities above, 
one of these model cases immediately fails the asymptotic freedom condition for $SU(10)$ even without any contribution from the Higgs sector which would only make things worse; in this 
regard, we need not realistically consider case ($b$) any further. As above, this is the result of requiring three copies 
of a single set of somewhat high-dimensional representations as is needed to reproduce the three SM fermion generations. We must also be especially mindful of both cases $(c)$ and ($e$) 
that come rather close in this regard so some additional care is necessary when examining the content of their required 
Higgs scalar symmetry breaking sectors as this may also lead to their exclusion for similar reasons. 

As before, to go any further, we must consider how the various and numerous fermion mass terms that may be needed in this $SU(10)$ scenario will be obtained from the general Yukawa 
coupling terms resulting from multiple (sub-)combination of factors:
\begin{equation}
\sim \bf {\bar {10}\cdot 45\cdot \bar{10}_H}+\bf {\bar {10}}\cdot 120 \cdot \bar{45}_H +\bf{45 \cdot 45 \cdot \bar{210}_H}+\bf{120 \cdot 120 \cdot 210_H}+...+\rm{perms.}+\rm{h.c.}\,.
\end{equation}
The potential contribution of the various Higgs scalar representations we find in this expression (together with the usual adjoint) to the $SU(10)$ $\beta$-function is given by the sum  
$(10+n_{10}+8n_{45}+28n_{120}+56n_{210})/6$ from which we see that, \eg, $n_{10}=n_{45}=n_{210}\geq1$, which is the common expectation,  already yields a contribution of 
$\delta \beta \geq 75/6$.  Given this, we can now also exclude both cases ($c$) and ($e$) immediately from further consideration since they will also violate the bound supplied by the 
asymptotic freedom 
constraint due to even this minimal Higgs scalar breaking sector. Before we turn to the remaining cases, it is interesting to note that in cases $(b)$ and $(c)$, where the SM $\bf{\bar 5}$'s 
are not embedded simply into $SU(10)$ $\bf{\bar{10}}$'s, one again finds that the Higgs fields responsible for generating the SM $d,e$-type masses are in $SU(5)'$ singlets thus excluding any 
contribution of $Q_{10}$ to $Q_D$ in these scenarios as usual.

We can easily make short work of case ($a$) as it falls into the same familiar pattern that we observed earlier for the $(a)$ cases of both $SU(8)$ and $SU(9)$ above; as before it is sufficient 
for our purposes to consider a single SM generation. Also as before, the realization of the SM fermions masses goes through without any issues (since the SM fermions lie in $SU(5)'$ singlet 
representations) via the now familiar couplings  
\begin{equation}
\sim \bf {(\bar 5,1)_{-1}}~\bf{(10,1)_2}~[\bf{(\bar 5,1)}_{-1}]_H + \bf {(10,1)_2}~\bf{(10,1)_2}~[\bf{(5,1)}_{-4}]_H+\rm{h.c.}\,,
\end{equation}
after selecting out one linear combination of the six $\bf {(\bar 5,1)_{-1}}$'s to play the role of the SM $\bf{\bar 5}$. As expected, however, the Yukawa term generating masses for the PM in this  
scenario is found to be inadequate to the task, \ie, with the familiar structure 
\begin{equation}
y_{ia}~[\bf {(\bar 5,1)_{-1}}]_i~\bf{(5,5)_{0}}~[\bf{(1,\bar 5)}_1]^a_H +\rm{h.c.}\,,
\end{equation}
where the $y$'s are Yukawa couplings as before, the index $i$ runs over the five remaining $\bf{(\bar 5,1)_{-1}}$'s and where, in principle, we can also allow for multiple Higgs scalars each with 
three $Q_D=0$ vevs that can all be different. We correspondingly recall, however, that there are only three distinct linearly independent sets of elements/directions in the group/field space for 
these multiple sets of scalar vevs (thus breaking $SU(5)' \to SU(2)'$) to maintain an unbroken $U(1)_D$. This leaves us with two of the five potential PM $SU(5)$ $\bf 5$ fermion fields 
remaining massless at the desired stage of symmetry breaking. As before, adding further additional Higgs scalars \ie, beyond three, is redundant and of no help due to the required limitation 
on the number of independent sets of vevs since there are only $Q_D=0$ directions are allowed in field space. {\it Without} this restriction, of course, it would be straightforward to generate the 
masses of all five these vector-like fermion PM representations. This eliminates the set of $n_g=1$ models from any further consideration and so we must turn our attention to the one 
remaining case ($d$).

In model ($d$), which is an $n_g=3$ scenario, both SM as well as PM fermions are found to lie in some common representations of the $55'1_{10}$ subgroup and many of the aspects of 
its symmetry breaking and mass generation aspects will be familiar from the previously examined cases above, albeit now with a bit more complexity. We begin by reconsidering the set 
of Yukawa couplings that can produce the masses for the various fermions which is quite rich in this particular scenario since all of the possible Higgs scalars are involved and some of 
the SM fermion fields now can lie in a $\bf{\bar {45}}$ instead of a $\bf{45}$ as they did in case ($a$):
\begin{equation}
\sim \bf {\bar {10}}\cdot(\bar {10}\cdot 45_H+\bar{45}\cdot 120_H +120\cdot \bar{45}_H)+\bar{45}\cdot(\bar{45}\cdot 210_H+120\cdot \bar{10}_H) +120\cdot 120 \cdot 210_H+\rm{h.c.}\,.
\end{equation}
This being the case, we are reminded that the full $SU(10)$ one-loop $\beta$-function for this model is given by $\beta_d=-62/3+(10+n_{10}+8n_{45}+28n_{120}+56n_{210})/6$ where 
the first term is the sum of the fixed gauge and fermion contributions. Note that the number of the larger Higgs scalar representations, $\bf{120, 210}$, is therefore quite restricted here.
We further observe that if we analyze the 245, two-component fermion degrees of freedom in case ($d$), we see that, in addition to the usual 45 SM fields, 60 of the them are purely dark 
sector fields while the remaining 140 are (hopefully) to be identified with 70, four-component vector-like PM fermions, all of which must obtain masses significantly in excess of the 
electroweak scale via suitable Higgs choices. The mass requirements for the purely dark sector fields are not as strict except that we'd like to identify the lightest such field having 
$Q_D\neq 0$ with DM and so it should have a mass roughly of order the $U(1)_D$ breaking scale. Of course, a purely dark scalar is also just as likely, and may even be favored, to be the 
light DM when all of this model's details and constraints are fully examined
 
Due to the complexities of the relevant couplings in this scenario, we will consider the generation of the fermion masses in the different sectors in some further detail here. 
In this model,  we can consider two quite distinct ways of embedding the SM and PM fermions into the various $SU(10)$ representations. The first, more obvious and more familiar 
approach, [here called subcase ($\alpha$)], is where three linear combinations of the eight $\bf {\bar 5}$'s amongst the eight $\bf {\bar {10}}$'s of $SU(10)$ 
are identified as the familiar SM fields. In a somewhat more complex embedding,  the SM fields can instead [termed subcase ($\beta$)] correspond to the three $Q_D=0$ members of the 
$\bf{(\bar 5,\bar 5)}_0$ in the $\bf{\bar {45}}$.  
In subcase ($\alpha$), since at least some of the SM fields are to be identified with $SU(5)'$ singlets, it is imperative that $Q_D$ cannot receive any contribution from $Q_{10}$. 
In either of these subcases, the three familiar SM $SU(5)$ $\bf {10}$'s can only be identified with the three $Q_D=0$ members of the $\bf{(10,5)}_{1}$ that lies in the $\bf {120}$ while the 
other two members of this representation must pair up with the $\bf{(10,1)}_2$ in the $\bf{\bar{45}}$ and the $\bf{(\bar{10},1)}_{3}$ in the $\bf{120}$ to obtain a total of 20 vector-like 
masses for the resulting 4-component fermions and then be identified with some of the PM.  Failure at this step would immediately exclude this model independently of the choice of subcase. 

Both of these possible 
scenarios have some rather non-trivial mass generation requirements but, as is immediately apparent, subcase ($\alpha$) is somewhat more straightforward. For either of these possibilities, 
at the $55'1_{10}$ level, one finds that 
there are 22 allowed Yukawa couplings: 4 for the purely PM sector involving fields which transform non-trivially under both $SU(5)$'s, 5 which solely concern dark sector fields with only 
non-trivial $SU(5)'$ transformation properties (both of which we can imagine obtaining masses at multi-TeV scales) and 13 which result in, \eg,  the breaking of the SM gauge symmetry 
and the generation SM fermion masses. Interestingly, also among the latter, are many `extra'  terms which are responsible for admixtures between the fields in the various sectors 
in addition to those producing the 
SM fermion masses themselves. This makes the overall fermion mass generation problem difficult to analyze in this particular setup partially due to, \eg, the previously discussed 
`left over' $\bf{\bar 5}$'s after the SM fields have obtain their masses. However, since the $SU(5)'$ and electroweak breaking scales are very different we may, at least approximately, treat them 
independently and roughly decoupled. Note that, as before, since either fields that will be identified as SM fermions or as $SU(2)_L$-breaking Higgs are to be found 
in $SU(5)'$ singlets in either of these subcases then $Q_D$ cannot have a contribution from $Q_{10}$ as is the norm. 

For either of these subcases, however, in the PM sector we must generate the vector-like fermion masses for the remaining two non-SM elements of the $\bf{10}$ representation (which are 
not given any electroweak masses and to which we will return below) by coupling them to the two different $\bf{\bar {10}}$'s as noted by employing the $55'1_{10}$-level mass terms, \ie,  
\begin{equation}
\rm{y_{-2}}~\bf{(\bar {10},1)_{-2}}~ \bf{(10,5)_1}~[\bf{(1,\bar 5)_1]_H} +\rm{y_3}~\bf{(\bar{10},1)_3}~ \bf{(10,5)_1}~[\bf{(1,\bar 5)_{-4}]_H}+\rm{h.c.}\,,
\end{equation}
with the coefficients $\rm{y_{-2,3}}$ now being the two relevant Yukawa couplings corresponding to these distinct $\bf {\bar{10}}$'s. The vevs of the two scalar fields, $[\bf{(1,\bar 5)_{1,-4}}]_H$, 
which are responsible for these mass terms, essentially consist of two distinct sets of (the $SU(5)'$ subgroup) $SU(3)'$, $Q_D=0$, antitriplets, each with the 3 non-zero vevs, \ie,  
$\sim v_i,v'_i,~i=1-3$, with $i$ being an $SU(3)'$ index. Thus, working in the $SU(5)'$ subspace and denoting the fields there as {\it upper case and primed} versions of their familiar 
SM $SU(5)$ analogs and with the two corresponding $SU(5)'$ singlets 
denoted by $S'^c_{-2,3}$, respectively, only one fermion bilinear arises from each of these two coupling terms, \ie, $\sim S'^c_{-2,(3)}D'_iv_i(v'_i)$~{\footnote {Repeated indices 
are summed over as usual.}}.  So, now going back to the full 
$55'1_{10}$ space, we observe that each term in the equation above leads to a different set of 10 degenerate PM masses, `eating up'  two of the five $\bf{10}$ fields in 
the $\bf{(10,5)_1}$ and leaving us the three sets of fields that are necessary to form the corresponding masses for the three generations of the $u$-type SM fermions. 
This Yukawa structure thus accounts for ($2\times10=$)20, out of the total 70 needed PM mass terms; so far, so good.

To go further, we need to choose between the two subcases, ($\alpha,\beta$), and this can be most easily done by examining the $55'1_{10}$-level mass term directly coupling five sets of 
intended SM $\bf{5}$'s to the corresponding $\bf{\bar 5}$'s, \ie,  
\begin{equation}
\tau~\bf {(\bar 5,\bar 5)_0} ~\bf{(10,5)_1}~[\bf{(\bar 5,1)}_{-1}]_H +\rm{h.c.}\,,
\end{equation}
where here $\tau$ is just another Yukawa coupling parameter. Analyzing this in the $SU(3)'$ subspace as we did above and employing the same suggestive notation, we see that 5 field 
bilinears can be formed
\begin{equation}
~~~~~~ \sim \epsilon_{ijk}(D^c_i)'(U^c_j)'v_k, ~~ \sim N' U_i'v_i, ~~ \sim E' D_i'v_i\,,
\end{equation}
where $i,j,k=1-3$ are $SU(3)'$ indices, so that going back to the full $55'1_{10}$ space we see that ($5\times 5=$)25 of the additional PM mass terms (out of the total of 70 required) 
can be generated in this way. Note that since, 
apart from the vevs, the only freedom here is the overall Yukawa, $\tau$, this construction directly leads to 25 degenerate PM mass terms. Now if we were in subcase ($\beta$), we would 
require {\it only} 10 mass terms out of this set as the 
other fields in the $\bar 5$ would need to be identified with the SM ones and this does not naturally happen here unless we reduce the number of non-zero vevs in the Higgs field - but this is 
something we can't to do as they are already needed to generate masses elsewhere. Given this, it appears that in this setup, the subcase ($\beta$) is excluded and this leaves us 
only with subcase ($\alpha$) which we will now assume is realized in what follows.

At this point one might be somewhat concerned about the SM particle masses themselves; we note that at the SM $SU(2)_L$-breaking level, this scenario will not only (hopefully) generate 
all the required SM fermion mass terms but will {\it also} lead to a rather complex mixing between the SM, PM and dark sector fields as well. If we were to consider the three generations of 
the SM fermion masses in {\it isolation} from these considerations (which is sufficient for the present discussion), in subcase ($\alpha$), the relevant Yukawa coupling terms are given by
\begin{equation}
y_{\sigma a} [\bf {(\bar 5,1)_{-1}}]^\sigma~\bf{(10,5)_1}~[\bf{(\bar 5,\bar 5)}_0]^a_H +\rho \bf{(10,5)_1}~\bf{(10,5)_1}~\bf{(\bar 5,\bar {10})_{-2}} +\rm{h.c.}\,,
\end{equation}
with $y_{\sigma a},\rho$ being Yukawa parameters and, as above, we have considered the possibility of additional scalar fields, labelled by the index $a$ as above, for the different 
SM $\bf{\bar 5}$ fermions as labeled by $\sigma$.  For convenience, we can label the corresponding Higgs vevs of the $\bf{(\bar 5,\bar 5)}_0]^a_H$'s as $w_{ai}, i=1-3$, similar to the 
above (although remember that here these are now $SU(2)_L$-violating vevs). 
Let's consider the second term first which links two identical SM $\bf{10}$'s to the corresponding $\bf{\bar 5_H}$. In the $SU(5)'$ subspace, we can denote the four allowed 
$SU(2)_L$-breaking, $Q_D=0$ vevs of the  $\bf{(\bar 5,\bar {10})_{-2}}$ as $u_{i=1-3},u_4$. Due to the identical nature of the fermions, this leads to only three non-trivial bilinears: 
$\sim \epsilon_{ijk}(D^c_i)'(D^c_j)'u_k$ with three different vevs. Projecting this back into the full $55'1_{10}$ space and scaling by $\rho$, we see that this will correspond to the three 
non-degenerate SM 
up-type quark masses that are required - another success for this model. Turning to the first term in the above expression, we recall that we wish to identify three linear combinations of the 
${(\bar 5,1)_{-1}}$'s with those of the SM and, further recalling that the $[\bf{(\bar 5,\bar 5)}_0]_H$ Higgs scalar arises from the $\bf{\bar{45}_H}$, we again find it useful to consider the 
coupling structure solely in the $SU(5)'$ subspace, here denoting the set of $SU(5)'$ singlets in the $[\bf {(\bar 5,1)_{-1}}]^a$'s simply as $S^c_a$. We then find that the only bilinears we can 
construct are 
of the form $\sim S^c_a(D_i)'w_{ai}$, \ie, {\it one} for each value of $a$; allowing for $a=1-3$ then produces the distinct masses for the three generations of $d,e$-type SM fermions. Now this 
implementation requires that $n_{45}=3$; assuming that $n_{10}=n_{120}=n_{210}=1$ still holds then this yields $\beta_d=-5/6$ so our asymptotic freedom constraint still remains satisfied. The 
generation of three distinct masses for the three generations of SM fermions is another success of this model.

Outside of the purely dark sector fields, there now remains only the issue of the five unpaired SM $SU(5)$ $\bf {\bar 5}$ fermions that we've already encountered for which we must generate 
vector-like masses so that they can also be identified as PM; these are the last of the ($70-20-25=$)25 mass terms that we noted above. In general, these couplings will take the form
\begin{equation}
\rm{z_{ab}}~\bf{(\bar 5,1)^a_{-1}}~ \bf{(5,10)_{-1}}~[\bf{(1,\bar {10})_2]_H}^b +\rm{h.c.}\,,
\end{equation}
with $z_{ab}$ being Yukawa couplings and in subcase ($\alpha$), which we are considering,  the index $a=1-5$, just labels these remaining unaccounted for $\bf{\bar 5}$'s while $b$ 
runs over the set of possible Higgs scalars.  Recall that the $\bf{(1,\bar {10})_2]_H} $ originates from a $\bf{\bar {45}_H}$ representation that we've already introduced three of in the 
previous discussion to generate the three generations of $d,e$-like SM fermion masses. As above, 
we turn to the $SU(5)'$ subspace to do the accounting of the number of independent mass terms generated in the present case; we will denote for simplicity the four, $Q_D=0$ vevs of the 
$\bf{[(1,\bar {10})_2]_H}$ by $u_{i=1-3,4}$ but these are not to be confused with those employed above arising from a different a $SU(5)'$ $\bf 10$ representation. For a {\it single} 
$[\bf{(1,\bar {10})_2]_H}$ and a single $\bf{(\bar 5,1)_{-1}}$ in this subspace, one correspondingly finds only a single bilinear, \ie, the combination $S^c[(U^c_i)'u_i +(E^c)'u_4]$, which then 
would generalize to 
$z_{ab}S^c_a[(U^c_i)'u_{bi} +(E^c)'u_{b4}]$. Momentarily forgetting the AF/$\beta$-function constraint (which tells us $b\leq 3$), this leads to at most four linearly independent combinations of 
vevs and, thus, only $4\times 5=20$ additional vector-like fermion PM mass terms beyond those constructed above. Once the $\beta$-function constraint is imposted, this is reduced to 
only $3\times 5=15$ mass terms based on the three allowed and previously introduced $\bf{\bar {45}_H}$'s implying that, in SM $SU(5)$ language, two pairs of $\bf{5}+\bf{\bar 5}$ PM fields 
will remain massless at the scale of $SU(5)'$ breaking. It is important to recall that the absence of the necessary mass terms can be partly traced to the requirement that we cannot give 
vevs to the $SU(2)'$ doublets within the various $SU(5)'$ Higgs representations as they necessarily must carry $Q_D \neq 0$ and would result in the breaking of $U(1)_D$. Such Higgs 
fields are mandatory as, although the PM fermion fields are vector-like with respect to the SM, they are chiral under $SU(2)'$ gauge group as was discussed earlier. Thus, at least at tree-level,
these fields will remain massless even after $SU(2)'=SU(2)_D$ breaks down to $U(1)_D$ so, no matter what happens in the purely dark sector this model is now also excluded, thus 
highly disfavoring $SU(10)$ scenarios. It is important to note that this problem would still persist, although with a somewhat reduced severity, even if the AF requirement were to be relaxed 
although we were brought rather quite close to success in this scenario

\subsection{Possible Future Directions: Going Beyond $SU(10)$  and Other Options}

When $N$ is small, \eg, $N\leq7$, $G_{Dark}$ is clearly not large enough to embed a $U(1)_D$ in a successful manner. However, the previously studied models have shown us that as 
$8\leq N\leq 10$ increases, our other constraints can quickly become more difficult to satisfy due to the tensions between the ever rapidly 
growing dimensionality of the non-fundamental representations for both the fermion and Higgs scalar fields, the requirement of asymptotic freedom above the $SU(N)$ unification scale, and 
the need to generate all of the SM and non-SM fermion masses with a restricted number of vevs so that $U(1)_D$ remains unbroken. This later problem, however, was seen to be 
somewhat alleviated by the presence of new types of Higgs fields in the second rank anti-symmetric representation of $G_{Dark}$ once $N\geq 9$, although not sufficiently so as to provide 
for a successful 
result in the examined cases.  A significant obstacle in the construction of a viable model was observed to be that a $SU(2)'$ survives after the initial dark gauge group breaking and that 
$U(1)_D$ must lie (at least in the examples discussed above) wholly within this $SU(2)'$. When $N>10$, these conflicting requirements will make finding a potentially successful model 
even more 
difficult. For example, consider two scenarios in $SU(11)$: first, an $n_g=3$ model\cite{Chen:2021ovs} with a relatively low number of fermion degrees of freedom (\ie, having `only' 341) 
given by $\bf {6(\bar {11})+2(\bar {55})+165}$. This setup requires at least one Higgs scalar in each of the $\bf{\bar{11}, \bar {55},165,330}$ and $\bf{462}$ representations to at least make at 
attempt at generating the necessary masses and so easily fails the test of asymptotic freedom. Similarly, a second, simpler $n_g=1$ scenario\cite{Vaughn:1979sm} in $SU(11)$, 
with only 351 degrees of freedom, 
$\bf {3[7(\bar{11})+55]}$ with a correspondingly simpler Higgs scalar sector, while possibly asymptotically free, now ends up following the well-known pattern of the case ($a$)-type models for  
$N=8,9,10$ discussed above and can be easily shown to not allow for all the required mass terms due to the remaining $SU(2)'$. Even this class of ($a$)-type scenarios, assuming the simplest 
of Higgs sectors, will also fail the asymptotic freedom constraint once $N>12$. Apparently, these multiple requirements, taken together, do not allow us to achieve our desired goal if the other 
model building assumptions above are not altered and apparently going to even larger values of $N$ will fail to buy much that is new.

While the asymptotic freedom requirement is clearly of some significant impact as we have seen, it is the fact that the $U(1)_D$ must lie totally within $SU(2)'$ that inhibits a larger 
number of scalar vevs from appearing, preventing a successful outcome{\footnote {As noted, at large values of $N$ this is alleviated to some extent by there also being distinct Higgs 
fields in the second-rank, antisymmetric representation of $SU(N-5)'$ thus increasing number of allowed $Q_D=0$ vevs.}}. 
This requirement, fundamentally, arose from the assumed gauge symmetry breaking structure of $SU(N)\to SU(5)\times SU(N-5)'\times U(1)_N$ and that $Q_D$ 
is found to be independent of $Q_N$, \ie, without any contributions arising from additional $U(1)$ factor and, in particular, that we can always choose a basis where $Q_D\sim T_3'$. In 
models where the relevant Higgs scalars are only in the (anti-)fundamental representation of $G_{Dark}$, this obstacle would require circumventing, or at least substantial softening, if we 
are to find an amiable solution. One possibility that we might imagine is to slightly modify the dark gauge group breaking 
pattern to allow for an extra $U(1)$ factor, thus providing additional freedom to embed $Q_D$. For example, one might consider beginning 
with the group $SU(N+1)$ selecting an model with a fixed representation content which is both anomaly and asymptotically free. Then we consider the breaking path 
$SU(N+1)\to SU(N)\times U(1)_{N+1} \to SU(5)\times SU(N-5)'\times U(1)_N \times U(1)_{N+1}$ which apparently seems to provided us with an additional $U(1)$ factor. 
To see if such an approach can be remotely viable, let us consider a pair of examples, both based on $SU(9)$ models previously examined above: ($i$) case ($a$)  and ($ii$) case ($e$). 

In example ($i$), we consider a single generation consisting of the anomaly free set of $5(\bf{\bar 9})+36$ representations which now under the breaking $SU(8)\times U(1)_9$ is just 
$[4(\bf{\bar {8}_{-1})+28_2]+[\bar{8}_{-1}+8_{-7}]+1_8}$ where the set of representations in the first bracket we recognize as a single generation of the $SU(8)$ case ($a$) (albeit now with a 
set of additional $U(1)$ quantum number attached) while the second bracket, with respect to $SU(8)$, is further pair of PM fields that are {\it not} present in the version of this now $SU(8)$ 
model previously considered above. Completing the picture, an additional $SU(8)$ singlet field is now also seen to be present. In comparison to the $SU(8)$ model
analysis above, we've gained the additional wanted freedom associated with the new $U(1)$ factor but, simultaneously, we have needed to add some 
new fermion fields whose masses we must also unfortunately now generate. This supplies  
new constraints that need to be satisfied in addition to those already encountered for the SM and PM fields examined previously. Recall that, in general, we now find that 
$Q_D=a_1\lambda_3+a_2 \lambda_8+bQ_8+cQ_9$, where $Q_{8,9}$ are just the $U(1)_{8,9}$ quantum number assignments of the various representations. 
Some algebra tells us that while the condition $c=-3b$ will allow for the generation of both the 
$\bf{\bar 5}\cdot \bf{10}$ and $\bf 10 \cdot 10$  SM particle masses, the corresponding requirements for all of the PM fields (including these newly introduced fields) cannot be met simultaneously   
even though the mass terms for the SM $\bf {\bar 5}$ and $\bf 10$ fields yield identical constraints on both of the $U(1)$'s individual contributions to $U(1)_D$. In this example, we've gained 
no ground by having this additional $U(1)$ factor due to the presence of the additional PM. 

Perhaps, if we could remove the additional constraints arising from the new PM states then 
greater success would become possible. To that end, we next, in example ($ii$), revisit $SU(9)$ case $(e$) wherein the three generations of SM fermions are assigned as 
$9(\bf {\bar 9})+84$, \ie, an example of an $n_g=3$ scenario. Again going to $SU(8)\times U(1)_9$ level, this corresponds to the representations 
$9(\bf {\bar 8}_{-1} ) +28_{-6}+56_3$+$9(\bf1_8)$, which  appears quite similar to the previously examined $SU(8)$ case $(e$) apart from the extra singlets and the overall additional $U(1)$ 
factor as we wanted. In this case we see that there are {\it no} additional PM-like states resulting from this procedure which is just what we desired to achieve, apparently alleviating the additional 
constraint found in the previous example.  However, in this scenario, unlike in ($i$), the generation of both the $\bf{\bar 5}\cdot \bf{10}$ and $\bf 10 \cdot 10$ SM particle masses lead to 
{\it different} requirements on the values of the $b$ and $c$ parameters so that the {\it total} number of constraints in this setup is found to be the same as in $(i$). In fact, one finds that 
the generation of the SM masses requires $b=c=0$ so that we are led back to the (failing) analysis in the previous subsection. Again, we find that we've gained nothing by adding this additional 
new $U(1)$ factor. Indeed one finds by further analysis that there is nothing special about these two examples and that these result are, unfortunately, rather general.
Although several possible scenarios seem to be opened by this initial idea, the additional constraints are found to more than outweigh any gains associated with the additional $U(1)$ degrees of 
freedom so that this approach is seen to fail.

Another possible path that one can imagine is to give up on the identification $G_{SM}=SU(5)$ and assume a somewhat larger SM gauge group such as, \eg, $SU(6)$. In doing so, still 
beginning with $G=SU(N)$, one trivially finds that the rank of $G_{Dark}$ is now reduced to $G_{Dark}=SU(N-6)'\times U(1)_N'$ which one might think could be advantageous. However, 
in doing so for fixed $G$ this reduces the number of dark sector diagonal generators on which $Q_D$ can depend (\ie, the rank of $G_{Dark}$ is now smaller by unity), hence, the number of 
diagonal generators out of which $Q_D$ can be constructed is also reduced. This then implies that the number of possible vevs of the various Higgs fields having $Q_D=0$ to maintain an 
unbroken $U(1)_D$ is also reduced making it more difficult to simultaneously generate masses for the set of SM $Q_D=0$ fermions as well as those for the remaining additional PM 
fermion fields. Thus, if anything, for fixed $G$, one would like to {\it increase} the rank of $G_{Dark}$ or at least increase the number of its diagonal generators contributing to $Q_D$ to allow for 
more possible Higgs vevs to generate the needed fermion mass terms. 

In the current study we have considered models which are in either of the $n_g=1$ or $n_g=3$ classes; there is also the possibility, not entertained above, of `$n_g=2+1$', wherein, by 
construction, one of the generations is embedded into the $SU(N)$ group fermion representation structure asymmetrically from the other two. Such an occurrence, at least partially, 
already happens in some of the $n_g=3$ models, \eg, in $SU(7)$ model ($c$), where two of the SM $\bf{10}$'s lie in a $\bf{21}$ and the other in a $\bf{35}$ or vice-versa. In principle, such an 
approach could lead to some additional model building flexibility but, based upon what we have seen above,  this would most likely occur in setups wherein none of the SM fermions of 
the two 'common' generations lie in the (anti-)fundamental representation of $SU(N)$. Clearly, we would be most interested in scenarios which are not just `simple' deconstructions of previously 
examined $n_g=3$ models as we gain nothing by doing this, \eg, the $SU(7)$ $n_g=3$ model ($c$) is seen to be composed of two sets of representations as are occurring in model ($a$) 
plus a single set from model ($b$).  Similarly, we see that  the corresponding $n_g=3$ model ($d$) is observed to combine one set of representations from model ($a$) and two from model 
($b$)\cite{Chen:2022xge}. The fact that both of these two models failed to satisfy all of our constraints gives some indication that the general $n_g=21$ set of models will also not meet with 
much success. However, a detailed study of this possibility in more realistic scenarios lies beyond the scope of the present work. 

In the analysis above, we have concentrated on the generation of the Dirac masses for the charged SM and PM fields; of course, if we were to find a successful scenario we would want 
to explore how the light neutrino masses might be generated and the nature of the associated phenomenology. Clearly, once we go beyond the standard $SU(5)$ the opportunities are many to 
generate interesting neutrino mass and mixing structures due to the existence of new neutral heavy vector-like leptons (some of which also may carry dark charges and can be identified as PM) 
which transform as SM singlets or as parts of one or more sets of ${\bf 5}+\bf{\bar 5}$'s under $SU(5)$ which are lie in SM isodoublets. For example, even in the simplest case of $SU(6)$, one 
sees that a single family will 
contains an additional such set of ${\bf 5}+\bf{\bar 5}$'s as well as two additional SM singlet fermions which open many possibilities for interesting scenarios\cite{Chacko:2020tbu}. Of course,  
the present model building structure will add extra complexities to the usual approaches as the $U(1)_D$ subgroup needs to remain unbroken down to $\lsim 1$ GeV and this may have 
a significant influence on the resulting neutrino mass spectra. However, to pursue this matter further we would first need a model that successfully generated all of the SM and PM masses and it 
might be more useful to first explore some of these possibilities from a bottom-up perspective.

There are, of course, many other potential directions and modifications to the approach followed above that one might try to explore to alleviate the problems we encountered some of which 
we will reserve for future work.

\section{Discussion and Conclusions}

The abelian KM portal model wherein a new gauge boson, the dark photon, associated with the gauge group $U(1)_D$, kinetically mixes with the hypercharge gauge boson of the SM 
offers a very attractive and well studied scenario for thermal DM at the $\lsim 1$ GeV mass scale. This KM is generated by loops of a set of portal matter fields which carry both dark and 
SM quantum numbers and should have masses above the electroweak scale. Specifically, fermionic PM must be vector-like with respect to the SM if it is to satisfy constraints arising 
from electroweak precision measurements, unitarity, and direct collider searches as well as the values of the Higgs boson loop-induced $gg,\gamma \gamma$ partial widths. It is natural to ask 
how the physics of the SM, dark and PM sectors might be combined into a single unified  
framework. Bottom-up approaches\cite{Rueter:2019wdf,Wojcik:2020wgm} in our class of models seem to indicate that an important first step in in this direction is to embed the 
$U(1)_D$ group into a non-abelian structure at a mass scale similar to or perhaps not too far above that associated with the PM fields themselves. An obvious issue is that we unfortunately 
lack a sufficiently large enough set of examples of this idea that 
would help guide us in a top-down approach. With this limitation in mind, in this paper we have attempted to construct, employing such a top-down approach, a unified scenario that combines the 
SM, dark and PM sectors into an overarching framework based on the assumption that all of the relevant gauge forces can be unified within a single $G=SU(N)$ gauge group. It was 
further assumed that $G$ then  
decomposes into the product $G_{SM}\times G_{Dark}$ with, for simplicity, the familiar $SU(5)$ playing the role of a stand-in for $G_{SM}$ with $U(1)_D$ assumed to be a diagonal 
subgroup of $G_{Dark}$.

Given this overall structural assumption, our model building was subsequently subjected to a rather large set of requirements, each of which - on its own - seems to be quite reasonable but 
which used collectively led to interesting and powerful constraints on the overall nature of the resulting model. For example, when these model building constraints were combined it was found to 
be impossible to generate all of the necessary electroweak scale or above mass terms for both the fermionic SM and PM fields while also simultaneously allowing a $U(1)_D$, under which 
the SM fields were singlets,  to remain unbroken far below the electroweak scale as required. However, we were able to get reasonably close to a satisfactory 
solution in the case of one of the $G=SU(10)$-based setups, especially so if the rather powerful $SU(N)$ asymptotic freedom requirement were to be dropped. The cases with larger values 
of $N$, though generally more constrained by the AF requirements, allowed for $G_{Dark}$ breaking and PM mass generation not only via the Higgs fields in the (anti-)fundamental 
representations of this group but also by second rank anti-symmetric tensor representations as well which provided an additional source of the needed scalar vevs. Quite generally this 
difficulty mostly resulted from the constraint, which followed directly from the model building requirements, that the $U(1)_D$ must 
solely originate from the unbroken diagonal subgroup of an $SU(2)'$ at the next to final stage of the symmetry breaking for $G_{Dark}$ and the assumption that $G_{Dark}$ itself had the 
largest rank possible given that $G_{SM}=SU(5)$ was assumed. This problem is then further accentuated by the inability of $Q_D$ to obtain a dependence of the $U(1)_N$ charge, $Q_N$. 
It would appear that more work on the bottom-up approach would also be useful to help clarify the physics at 
the intermediate breaking scale as well as the varieties of possible group representations of $G$ that can be shared by both SM and PM fermion fields

Several possible hopeful directions to overcome these model building difficulties were discussed and identified and will become the subject of future work.

%------------------------------------ ACKNOWLEDGEMENTS ---------------------------------------%
\section*{Acknowledgements}
The author would like to particularly thank J.L. Hewett, D. Rueter and G. Wojcik for valuable discussions related to the early aspects of this work.  This work was supported by the 
Department of Energy, Contract DE-AC02-76SF00515.

%------------------------------------------- REFERENCES -------------------------------------------%

%-------------------------------------------------- END --------------------------------------------------%

\end{document}